\documentclass[fleqn,usenatbib]{mnras}

\usepackage[T1]{fontenc}

\DeclareRobustCommand{\VAN}[3]{#2}
\let\VANthebibliography\thebibliography
\def\thebibliography{\DeclareRobustCommand{\VAN}[3]{##3}\VANthebibliography}

\usepackage{graphicx}	
\usepackage{amsmath,amsfonts,mathabx}	

\usepackage{fix-cm}
\usepackage{pdfpages}
\usepackage{siunitx} 
\usepackage{hyperref}
\usepackage{txfonts}
\usepackage{listings}
\usepackage{booktabs}	
\usepackage{subfig}

\DeclareSIUnit\angstrom{\text{\AA}}

\title[Unified properties of SMBH winds in RQ and RL AGN]{Unified properties of supermassive black hole winds in radio-quiet and radio-loud AGN}

\author[S. Mestici et al.]{
S. Mestici,$^{1,2}$
F. Tombesi,$^{1,3,4}$
M. Gaspari$^{5}$
E. Piconcelli$^{3}$
and F. Panessa$^{6}$
\\
$^{1}$Physics Department, Tor Vergata University of Rome, Via della Ricerca Scientifica 1, 00133 Rome, Italy\\
$^{2}$Physics Department, Sapienza University of Rome, Piazzale Aldo Moro, 00185 Rome, Italy\\
$^{3}$INAF – Astronomical Observatory of Rome, Via Frascati 33, 00040 Monte Porzio Catone, Italy\\
$^{4}$INFN - Rome Tor Vergata, Via della Ricerca Scientifica 1, 00133 Rome, Italy\\
$^{5}$Department of Physics, Informatics and Mathematics, University of Modena and Reggio Emilia, 41125 Modena, Italy\\
$^{6}$INAF – Istituto di Astrofisica e Planetologia Spaziali, via Fosso del Cavaliere 100, I-00133 Roma, Italy}

\date{Accepted 2024 June 27. Received 2024 June 10; in original form 2024 February 01}

\pubyear{2023}

\begin{document}
\label{firstpage}
\pagerange{\pageref{firstpage}--\pageref{lastpage}}
\maketitle

\begin{abstract}
Powerful supermassive black hole (SMBH) winds in the form of ultra-fast outflows (UFOs) are detected in the X-ray spectra of several active galactic nuclei (AGN) seemingly independently of their radio classification between radio quiet (RQ) and radio loud (RL). In this work we explore the physical parameters of SMBH winds through a uniform analysis of a sample of X-ray bright RQ and RL AGN. We explored several correlations between different wind parameters and with respect to the AGN bolometric and Eddington luminosities. Our analysis shows that SMBH winds are not only a common trait of both AGN classes but also that they are most likely produced by the same physical mechanism. Consequently, we find that SMBH winds do not follow the radio-loudness dichotomy seen in jets. On average, a comparable amount of material accreted by the SMBH is ejected through such winds. The average wind power corresponds to about 3 per cent of the Eddington luminosity, confirming that they can drive AGN feedback. Moreover, the most energetic outflows are found in the most luminous sources. We find a possible positive correlation of the wind energetics, renormalized to the Eddington limit, with respect to $\lambda_{Edd}$, consistent with the correlation found with bolometric luminosity. We also observe a possible positive correlation between the energetics of the outflow and the X-ray radio-loudness parameter. In general, these results suggest an underlying relation between the acceleration mechanisms of accretion disc winds and jets.

\end{abstract}




\section{Introduction}\label{intro}

Active Galactic Nuclei (AGNs) are extremely luminous astrophysical objects (i.e. $L_{bol}$ up to $10^{48}$ erg/s) located at the centers of some galaxies, powered by the accretion of matter onto supermassive black holes (SMBH, i.e. $M_{BH} >$ $10^6 M_\odot$). The term AGN comprises a wide range of objects that were historically separated by their observational features regarding a specific electromagnetic wavelength (see review by \citet{Padovani17}). As more AGNs with intermediate characteristics are still being discovered, these classes are constantly being revised. Currently, this "zoology" is partly explained in the framework of the Unified Model (\citet{Antonucci93}, \citet{Urry95}), where the accretion rate and the orientation with respect to the line of sight of the disc are the variables that lead to different spectral features (\citet{Netzer15} and references therein).\\ A still poorly understood aspect is their duality in the radio band. AGNs are classified as Radio-Loud (RL) when the ratio of radio to optical emission (R), defined as the flux density at 5 GHz over the one at 2500 $\si{\angstrom}$, is R$\geq$10, whereas Radio-Quiet (RQ) when 0.1$\le$R$\le$1 (\citet{Kellermann89}). In RL AGNs the bulk of the emission is due to synchrotron radiation produced by a collimated relativistic jet, and present evidence suggests that its physical origin lies in the proximity of the central SMBH (see \citet{Blandford19}). Furthermore, it appears that the BH spin and the kinetic power of the jet are closely linked (\citet{Chen21}). On the other hand, RQ AGNs spectra are dominated by the accretion disc emission. However, there is an increasing evidence of a gradual distribution of AGN radio power, instead of a sharp division between the two classes.\\ In terms of their overall prevalence, RL AGNs are less common than RQ, accounting for approximately $10-15\%$ of the AGN population. This distribution seems to be related to the SMBH mass function (\citet{Graham07}) as a significant correlation is observed between the radio luminosity (L$_{R}$) and the M$_{BH}$ ($L_{R}\propto M_{BH}^{2.5}$ (\citet{Franceschini98}, \citet{McLure04}, \citet{Best05}) suggesting that strong radio emission is connected to AGNs with a greater M$_{BH}$. In support of these studies, \citet{Laor00} found that nearly all PG AGNs with $M_{BH} \geq 10^9$ M$_{\odot}$ are RL, while those with M$_{BH}\leq 3\times10^8$ M$_{\odot}$ are RQ.\\

Following the first detections of fast ionized outflows (\cite{Chartas02}, \citet{Pounds03} and \citet{Reeves03}), a growing amount of work has gone into looking for AGN outflows in various interstellar medium (ISM) phases in the last decade. At this time, winds are indeed detected at various distances from the AGN innermost regions and with distinct ionization states: (i) at sub-pc scales through the detection of blueshifted highly ionized Fe K-shell transitions with velocities $\sim$0.1 $c$ or even higher, i.e. Ultra-Fast Outflows (UFOs) (see \citet{King15}, \citet{Tombesi10}; (ii) at pc scales via warm absorbers (WA) and broad absorption lines (BAL) (\citet{King15}, \citet{Tombesi13}, \citet{Serafinelli19}, \citet{He19}, \citet{Vietri22}) and the blueshift of the C IV emission line (\citet{Gaskell82}, \citet{Vietri18}); (iii) at kpc scales through different gas phases such as ionized gas (i.e. [OIII] emission lines, \citet{Harrison12}, \citet{Cresci15}) and molecular gas (e.g \citet{Feruglio10}, \citet{Bischetti19}). An exhaustive review of the topic can be found in \citet{Laha21}. Theoretical models of AGN-driven outflows (e.g. \citet{Fau12}) suggest that, in an energy conservation scenario, the kinetic energy of nuclear fast outflow is transferred to the ISM and drives the kpc scale flows. Outflows are considered one of the fundamental mechanisms by which the central SMBH interacts with its host galaxy, providing an efficient tool to regulate star formation, cooling flows, and drive correlations between the $M_{\rm BH}$ and the host properties (\citet{Voit15,Gaspari19}). Direct observations of the interaction between ISM and AGN winds have been collected so far (\citet{Cano12} and references therein), revealing a spatial anti-correlation between outflows and actively star-forming regions.\\ Moreover, \citet{Fiore17} found strong correlations between $L_{bol}$ and the cold and ionized wind mass outflow rate and kinetic power, showing that in galaxies hosting powerful AGN driven winds, the depletion timescale and the molecular gas fraction are 3-10 times shorter and smaller than those of main-sequence galaxies with similar star-formation rate, stellar mass and redshift.\\ 

Systematic studies of the X-ray spectra for a sample of z$\leq$ 0.1 AGN revealed that about 40\% of them have highly ionized UFOs with average velocities between 0.1 and 0.3 c (e.g. \citet{Tombesi10}, \citet{Gofford13}). Radiation and magnetic driving, or more likely a combination of the two, are the mechanisms suggested to explain the acceleration and launching of the X-ray absorbing material to mild relativistic velocities (e.g. \citet{Fukumura10},\citet{Fukumura14}). In the first scenario, the gas initially rises upward from the disc and is then pushed outward radially, accelerated by radiation pressure (i.e. Compton scattering and/or UV line absorption). However, this acceleration mode requires luminous AGN ($L_{bol}$>0.1$L_{Edd}$) and the upper limit on $v_{out}$ is $\sim$ 0.2c. In magnetohydrodynamic driven models, an outflow can be released from the disc depending on the magnetic field configuration, i.e. when the angle between the poloidal component and the disc reaches a certain threshold. A fraction of the accreting plasma can then be launched with a quasi-Keplerian velocity profile and accelerated along the magnetic field lines (in the so-called `magnetic tower' effect). This mode requires an accretion disc that is strongly magnetized, which is similar to the relativistic jet's initial condition. 
More generally, we refer to such outflows as `micro winds', as they can also be generated without the presence of a magnetized disc, in different scenarios (e.g., via radiative feedback).\\ These outflow and radiation pressure can prevent further accretion onto the SMBH disrupting the inflow material and leading to a self-regulating mechanism. Moreover, as the possible trigger of multi-scale outflows (\citet{Fau12}), UFOs are the starting point to understand the feedback and feeding processes that characterize the AGN-host galaxy interaction (\citet{Gaspari20}).
A state-of-the-art theoretical scenario is that SMBH feeding and feedback are recursively shaped by Chaotic Cold Accretion (CCA; e.g., \citealt{Gaspari13,Gaspari17,Maccagni21,McKinley22,Olivares22}). In CCA, the cold gas condenses out of the galactic hot halo and recurrently rains onto the micro-scale AGN, triggering ultrafast outflows. Being self-similar, CCA is expected to occur regardless of radio activity or AGN classifications.\\

The main objectives of this work are to: (a) analyze the presence of UFOs in RL and RQ AGNs to better understand the underlying differences between the two classes of the same physical phenomenon and (b) characterize the physical parameters of UFOs and their correlation with the AGN bolometric luminosity, $L_{bol}$. A systematic search for X-ray UFOs has been reported currently only in a sample of local (z$\leq$0.2) RL AGNs (\citet{Tombesi14}). In order to perform a comparative statistical study with respect to RQ AGNs, here we need to consider only UFOs detected in local sources. For statistical studies of X-ray UFOs focused only on high-z RQ AGNs, we refer the reader to other recent works (e.g., \citet{Chartas21b}; \citet{Matzeu23}).\\
The model and assumptions made to infer the physical parameters of the disc winds are explained in Section 2, along with the description of the AGN sample and additional considerations on the possible sources of uncertainty. In Section 3 we provide a discussion of the result, and in Section 4 we summarize our conclusions. Throughout this paper, we assume a flat $\Lambda$CDM cosmology with ($\Omega_M$, $\Omega_\Lambda$) = (0.3,0.7) and a
Hubble constant of 70 km s$^{-1}$ Mpc$^{-1}$.

\section{Data Analysis} \label{data}

In this work, the physical parameters of the UFOs detected in \citet{Tombesi14}, i.e. a sample of RL AGN observed with $XMM$-$Newton$ and $Suzaku$, are thoroughly examined. Moreover, the same parameters are derived for the RQ AGN samples of \citet{Tombesi11}, \citet{Tombesi12} and \citet{Gofford15} using the same methods. These works in the literature are based on a large and well-selected sample of sources and allow us to homogeneously compare the disc winds properties of the two types of AGN and study possible correlations.\\ 
Information on $M_{BH}$, the AGN unabsorbed luminosity in the X-ray band ($L_{x}$), defined over a range E$=$2~keV -- 10~keV, and the outflow equivalent hydrogen column density ($N_{H}$), ionization parameter (log$\xi$) and the velocity ($v_{out}$) was used to obtain other crucial physical parameters, as explained below. 

\subsection{Outflow Parameters} \label{outflow}

The AGN ionizing luminosity $L_{ion}$, defined in an energy range E$=$13.6~eV -- 13.6~keV, was estimated for the RL sample of \citet{Tombesi14} using $L_x$ and assuming a typical power-law continuum emission with a photon index $\Gamma = 1.8$, which is a value commonly measured in the X-ray spectra of RL AGNs (see \citet{Nandra94} and references thereafter). Instead, the $L_{ion}$ of the AGNs in the RQ AGN samples of \citet{Tombesi11}, \citet{Tombesi12} and \citet{Gofford15} is directly gathered from the respective works.\\ As X-ray outflows are relatively compact, an upper limit on the line of sight (LOS) projected location can be derived from the definition of $\xi= L_{ion}/nr^2$ (\citet{Tarter69}), and requiring that the thickness of the absorber does not exceed its distance from the BH, $N_H \simeq n\Delta r < nr$ (e.g. \citet{Crenshaw12}):

\begin{equation}
    r_{max} = \frac{L_{ion}}{N_H\xi}
\end{equation}

We specify that in those cases where the tabulated wind properties have only lower limits, which is generally true for $\xi$ and $N_H$, the conservative limit was adopted in the computation of the outflow parameters. An estimate of the lower limit for the UFO distance from the BH can be derived considering the escape distance relative to the outflow velocity. The escape velocity at a distance $r$ is $v_{esc}=\sqrt{2GM_{BH}/r}$. In the approximation $v_{out}=v_{esc}$, i.e. the outflow velocity measured along the LOS is equal to the escape velocity at that distance, then:

\begin{equation}
r_{min}= \frac{2GM_{BH}}{v_{out}^2}\
\end{equation}

These equations allow us to estimate the lower and upper limits of the outflow location, although with relatively large uncertainties. Better constraints are currently limited by the quality of the data and available models, and thus this is a conservative approach to estimate the wind location (e.g.,\citet{Crenshaw12}; \citet{Tombesi12} and \citet{Tombesi13}; \citet{Gofford15}; \citet{Laha21}; \citet{Chartas21}).\\

The mass outflow rate, $\dot{M}_{out}$, is defined as the mass flux carried by the outflow and is a critical parameter to understand the energetics of these phenomena. $\dot{M}_{out}$ is related to the geometry of the system and, as such, requires modeling of the outflow structure that can only be approximated. The standard formula adopted in a thin spherically symmetric scenario is (\citet{Gofford15}):

\begin{equation}
\dot{M}_{out} = 4\pi C_g r^2 m_p n v_{out}\
\end{equation}

where the product $C_g = \Omega b\leq 1$ is called the "global filling factor" and accounts for both the fraction of the solid angle occupied by the outflow ($\Omega$) and how much of the volume is filled by the gas ($b$); $m_p$ is the proton's mass; $n$ and $v_{out}$ are the outflow density and velocity, which can be assumed constant for a thin shell. Due to its dependence on gas ionization and clumpiness, the estimate of $b$ is complex. At low intermediate ionization states, the flow is likely to be clumpy or filamentary. This is supported by CCA models (e.g. \citet{Gaspari13}) which produce an intrinsic chaotic clumpiness due to top-down multiphase condensation rain. In high ionization states, as in the case of UFOs, $b$ can be considered to be largely smooth. We adopt a $b\simeq 1$ and a mean value for $\Omega$ of $\Omega=0.4$, as can be estimated from the fraction of Fe K outflows observed in sample studies (see \citet{Tombesi10} and \citet{Gofford13}). Thus, the global filling factor is $C_g \simeq 40\%$.\\ Using the estimate of the lower and upper limits for the outflow distance, the mass outflow rate is given by:

\begin{equation}
\dot{M}_{out}^{max} = 4\pi C_g m_p L_{ion} \xi^{-1} v_{out}\
\end{equation}

\begin{equation}
\dot{M}_{out}^{min} = 8\pi C_g G M_{BH} m_p N_H v_{out}^{-1}\
\end{equation}

Assuming that the outflow has reached a steady terminal velocity by the point at which it is observed, the instantaneous mechanical power can be estimated as $L_{out}= (1/2) \dot{M}_{out}v_{out}^2$ (see also \citet{Gaspari17}). Therefore, substituting in the previous equation, the following relations can be simply obtained:

\begin{equation}
L_{out}^{max} = 2\pi C_g m_p L_{ion} \xi^{-1}v_{out}^3\
\end{equation}

\begin{equation}
L_{out}^{min}= 4\pi C_g G M_{BH}m_p N_H v_{out}\
\end{equation}

The rate at which the outflow transports momentum in the host galaxy environment is given by $\dot{p}_{out}=\dot{M}_{out}v_{out}$. This physical quantity can also be regarded as the force that the outflow exerts over the interstellar medium or the force that is required to accelerate the outflow to its current state:

\begin{equation}
\dot{p}_{out}^{max}= 4\pi C_g m_p L_{ion} \xi^{-1} v_{out}^2\
\end{equation}

\begin{equation}
\dot{p}_{out}^{min}= 8\pi C_g G M_{BH}m_p N_H\
\end{equation}

\subsection{Other parameters} \label{params}

In addition to characterization of the outflow physical properties, other intrinsic features of the AGNs were also derived. The values of the bolometric luminosity for each source are taken from the literature or, where absent, determined as $L_{bol}=k_{bol}\times L_x$, where $k_{bol}$ is the bolometric correction factor. The latter is defined for each source as:

\begin{equation}
    k_{bol}=a\times[1+(log(L_x/L_{\odot})/b)^c)]
\end{equation}

where a = 15.33$\pm$0.06, b = 11.48$\pm$0.01 and c = 16.20$\pm$0.16 (\citet{Duras20}). The reference list for each source is provided in Table \ref{table:2}. The momentum rate of AGN radiation is then obtained as $\dot{p}_{rad}= \frac{L_{bol}}{c}$, and the mass accretion rate on the SMBH is obtained as $\dot{M}_{acc}=\frac{L_{bol}}{\eta c^2}$ where a standard accretion efficiency of $\eta=0.1$ was considered. Lastly, the Eddington ratio is simply computed as $\lambda=\frac{L_{bol}}{L_{edd}}$.\\In order to obtain further insights into the correlation between different wavelengths and the properties of disc winds, we also compute the X-ray radio loudness R$_x$ of each source as the ratio between radio luminosity at 5 Ghz and $L_x$, i.e. \( R_x = L_R/L_x\). We collected from the literature the available radio fluxes at 1.4GHz and derived the flux at 5GHz as \(S_{5} = S_{1.4} \times 10^{0.7 \times log_{10}(5/1.4)}\). The radio luminosities and the respective references for the radio flux are reported in table 2 of Appendix B. The disc wind parameters are first derived in physical units and then normalized to the individual M$_{BH}$ as explained below:
\begin{itemize}
\item{The distances are converted in units of the Schwarzschild radius, where $r_S= \frac{2GM_{BH}}{c^2}$;}
\item{The mechanical power is normalized to the Eddington luminosity $L_{Edd}$, where $L_{Edd}\simeq1.3\times10^{38}(\frac{M_{BH}}{M_{\odot}})$ erg~s$^{-1}$;}
\item{The mass outflow rate is normalized to $\dot{M}_{acc}$;}
\item{The momentum rate of the outflow is normalized to $\dot{p}_{rad}$.}
\end{itemize}

Indeed, despite the fact that the observed sizes of AGNs, which are usually determined by $M_{BH}$, vary, they also appear to share a similar physical structure, as suggested by the Unified Model (\citet{Urry95}). In particular, a higher $L_{bol}$ often implies a higher $M_{BH}$ and this extreme radiation field is responsible for both the sweeping of the accreting gas, due to its radiation pressure, and dust sublimation. Therefore, it is intuitive to assume that, based on the bolometric luminosity, the same gaseous structure describing an AGN may be discovered at different distances from the center. Deriving scale-invariant physical parameters that can describe the system independently of $L_{bol}$, is essential to understand the phenomenology of the outflow and their relation to the AGN types. In our study, we manage this task by performing the aforementioned normalization of the outflow parameters.

\subsection{Description of the dataset} \label{descr}

The initial sample consists of the RQ and RL AGNs described in \citet{Tombesi12}, \citet{Tombesi14}, and \citet{Gofford15}. Among these observations, for each source, only X-ray outflows with velocities higher than $\simeq$1\% of the speed of light were selected. This threshold allows us to identify the outflows most likely to be launched from the accretion disc (e.g., \citet{Tombesi10} and \citet{Gaspari17}) and not from further distances. In Appendix A, the estimated physical parameters of UFOs are reported. From Table A.1 to Table A.5 are shown the values for the \citet{Tombesi14} sample; from Table A.6 to Table A.10 for \citet{Gofford15}; and from Table A.11 to Table A.15 for \citet{Tombesi12}.\\ Subsequently, the final values of the UFO parameters for each AGN were derived as an average between different observations of the same source. The results reported in Appendix B show the mean between the upper and lower limits of the UFO parameters, where the uncertainty is the range given by the previous formulae. The classification of sources in RQ and RL AGN was also highlighted. Table \ref{table:1} shows the mean values of the UFO parameters for the whole RL and RQ AGN sample investigated in this work. 

\subsection{Possible uncertainties and biases} \label{biases}
We underline again that, in the following analysis, the confidence range for each parameter is given as the interval between its maximum and minimum values. Moreover, the formulae shown in Sec.~\ref{outflow} consist mostly of physical limits and approximations.  One of the main issues when considering uncertainties is that $log\xi$, $N_H$ and $v_{out}$ may be subject to possible systematics and effects of sample selection. For instance, different assumptions on the velocity broadening of the lines and the gas elemental abundances can generate variations in the estimated $N_H$ (see \citet{Tombesi13}) par 4.3). Therefore, gathering results from analyses reported in many different papers is not recommended, as the diverse analysis methods and assumptions employed could increase the scatter in the obtained values. A physical factor to take into account is the possibility of intrinsic inhomogeneities and variability in the absorbers that are not described by the models mentioned above. Moreover, we also note that the radio flux at 1.4GHz, used to calculate $R_x$, is subject to uncertainties which depend on the area adopted for the source measurement.\\ 

Additional parameters that could contribute to intrinsic uncertainties are the angle with respect to the line of sight and the opening angle, although statistical studies mitigate this problem, showing a UFO detection rate of $\sim$ 40\% for both RQ and RL AGNs (e.g., \citet{Tombesi10}, \citet{Tombesi14}).\\ Finally, the reported RL AGNs with disc winds are classified as FRII radio sources, known to be X-ray brighter than FRIs (e.g.,\citet{Hardcastle09}). This is a selection effect due to the need for a higher signal-to-noise ratio to detect spectral features that allow us to probe only the X-ray brighter population of RL AGNs, which in turn has an effect on our sample $R_x$ distribution. Moreover, FRII are in general more active sources with greater accretion rate than FRI, therefore, UFOs are expected to be observed more in the former (\citet{Best12}.\\ 
Given the previously explained premises, we are still confident that our results are indicative of the main physical conditions of the population of sources here investigated.

\section{Discussion}

In this work, we present a systematic analysis of the physical parameters of UFOs detected in a large sample of AGN consisting of 27 sources, 7 of which are RL and 20 RQ. In the following, we develop our study assuming that the 7 RL AGNs that host UFOs are representative of the key characteristics of their class. The conclusions reached are collected in the tables in Appendix A and are discussed in the next paragraph. The analysis then concentrates on potential correlations between the average values of the outflow parameters given in the tables in Appendix B.

\subsection{Common origin for disc winds in X-ray bright RQ and RL AGNs} \label{origin}

We start with a comparison of the average parameters for every source in both AGN classes (see Tab. \ref{table:1}). The confidence range for each parameter is derived as the standard deviation of the measures.\\

\begin{table}
\caption{Mean values and error on the mean of the AGNs intrinsic properties and UFO parameters for RL and RQ AGN and p-value of the KS-test between the two distributions.} 
\label{table:1}      
\centering                      
\begin{tabular}{l c c r}        
\hline\hline      
Parameter  & RL & RQ & p-value \\    
\hline
log($M_{BH}$) & $8.3\pm0.6$ & $7.6\pm0.9$ & 0.16\\ 
log($L_{bol}$) & $45.4\pm0.5$ & $44.7\pm0.8$ & 0.01\\ 
log$\lambda_{Edd}$ & $-1.1\pm0.5$ & $1.0\pm0.5$ & 0.50\\ 
\hline  
log$\xi$ & $4.1\pm1.2$ & $4.1\pm0.6$ & 0.56\\
log$N_H$ & $22.7\pm0.6$ & $22.6\pm0.5$ & 0.75\\
$v_w/c$ & $0.16\pm0.10$ & $0.16\pm0.08$ & 0.99\\
log($r_{out}/r_{S}$) & $3.2\pm0.9$ & $2.9\pm0.6$ & 0.64\\ 
log($\dot{M}_{out}/\dot{M}_{acc}$) & $0.8\pm0.7$ & $0.4\pm0.6$ & 0.61\\
log($L_{out}/L_{Edd}$) & $-1.6\pm0.6$ & $-1.8\pm0.8$ & 0.77\\ 
log($p_{out}$/$p_{rad}$) & $0.8\pm0.6$ & $0.4\pm0.7$ & 0.74\\
\hline           
\end{tabular}
\end{table}

A statistical comparison between the parameter distributions of the two classes is carried out using the two-sample Kolmogorov–Smirnov test (Smirnov (1939)). The linear regression p-value for each independent variable tests the null hypothesis that the two populations are drawn from the same underlying distribution. In particular, the p-value identifies the point where the integral of the conditioned probability density of the variable reaches a certain value. Therefore, the p-value can be understood as the confidence level with which the null hypothesis is rejected in favor of the alternative. A standard confidence level for rejecting the null hypothesis is 95\%, i.e., if the p-value is less than 0.05. It should be noted that the p-value is suitable to confirm or deny the null hypothesis but offers no other information on the distributions in the latter case. The KS-test p-values for each outflow parameter are collected in Tab. \ref{table:1}\\

\begin{figure}
\centering
\includegraphics[width=0.75\columnwidth]{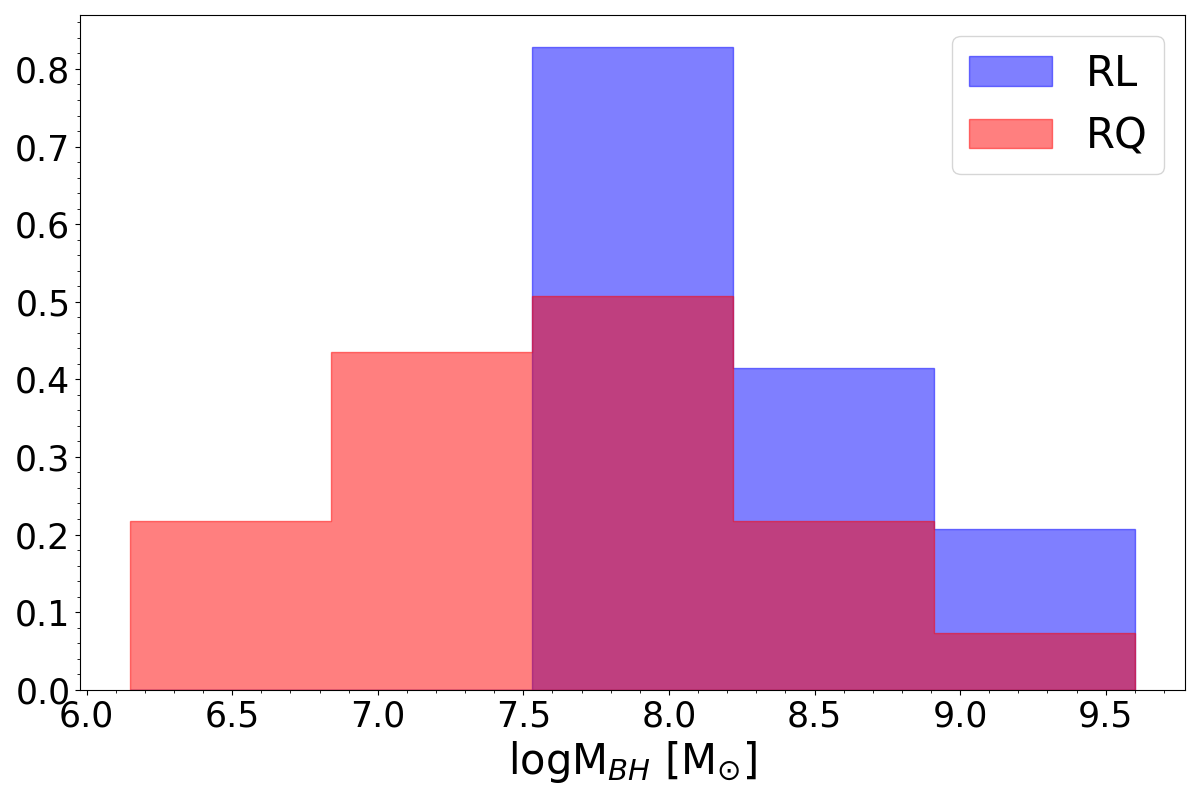}
\includegraphics[width=0.75\columnwidth]{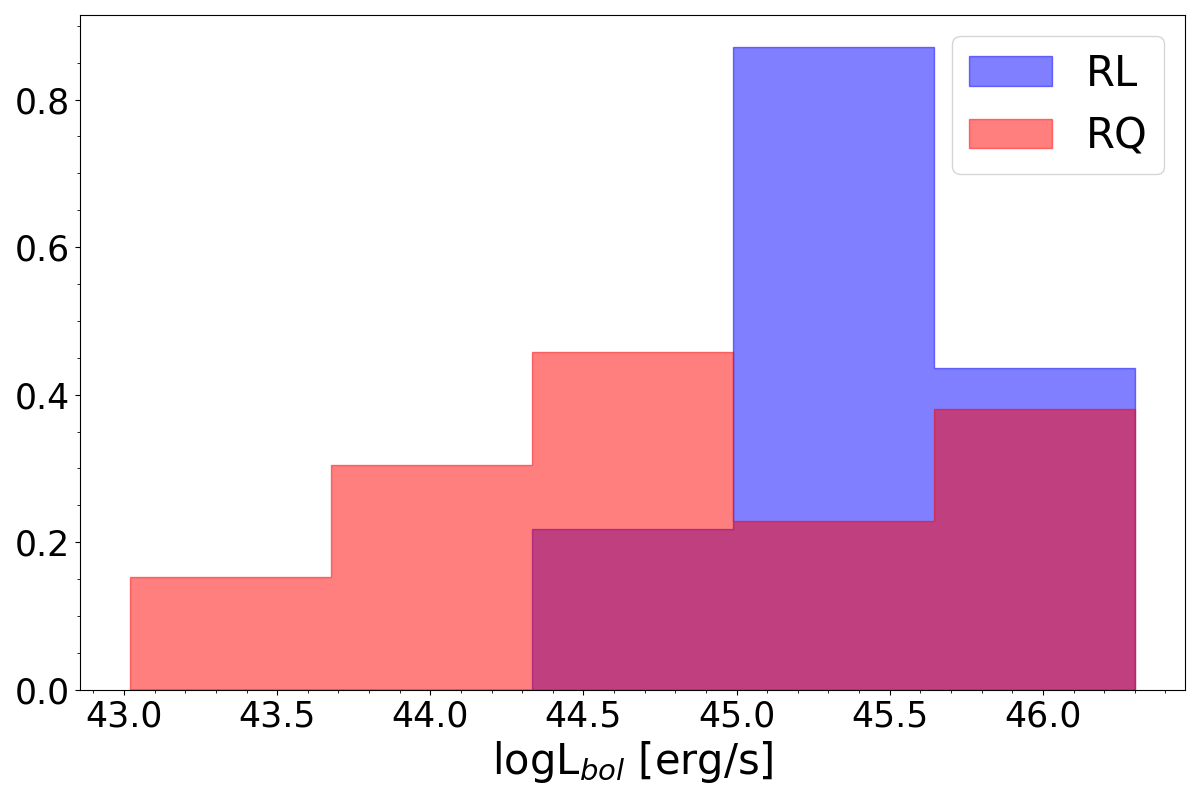}
\includegraphics[width=0.75\columnwidth]{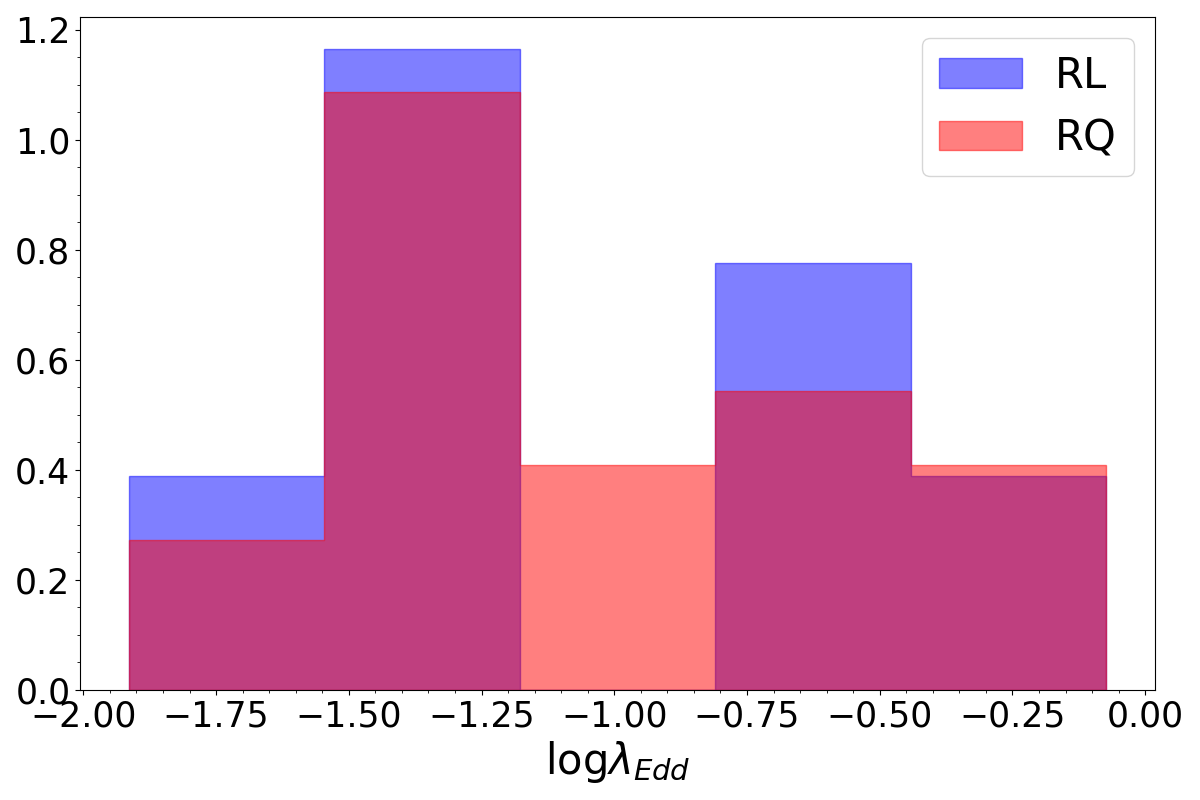}
\caption{Distributions of the logarithm of the SMBH mass in solar mass units (upper panel), bolometric luminosity (central panel), and Eddington ratios (lower panel) for the combined sample of RL (blue) and RQ (red) AGNs with detected outflows.}
\label{distrib1}
\end{figure}

A partial overlap can be observed by comparing the normalized distributions of $M_{BH}$, $L_{bol}$ and $\lambda_{Edd}$ for the two AGN classes (see Fig. \ref{distrib1}). In particular, RL sources are systematically clustered towards the higher-end of the parameter space, which is in line with the known trends due to a slightly larger $M_{BH}$. The KS-test rejects the null hypothesis only for $L_{bol}$ distributions with a p-value$=$0.01. Indeed, even if the means of the two distributions are consistent i.e. log($L_{bol}$)$_{RL}=45.4\pm0.5$ and log($L_{bol}$)$_{RQ}=44.7\pm0.8$, the RL distribution is more skewed and asymmetric. Although for $M_{BH}$ the null hypothesis cannot be rejected, the confidence level still lies at $\sim$84\%. These results underscore that there are differences in the intrinsic properties of two AGN populations. This point is crucial in the following analysis, where we compare the physical properties of the outflows.\\ 

\begin{figure}
\centering
\includegraphics[width=0.75\columnwidth]{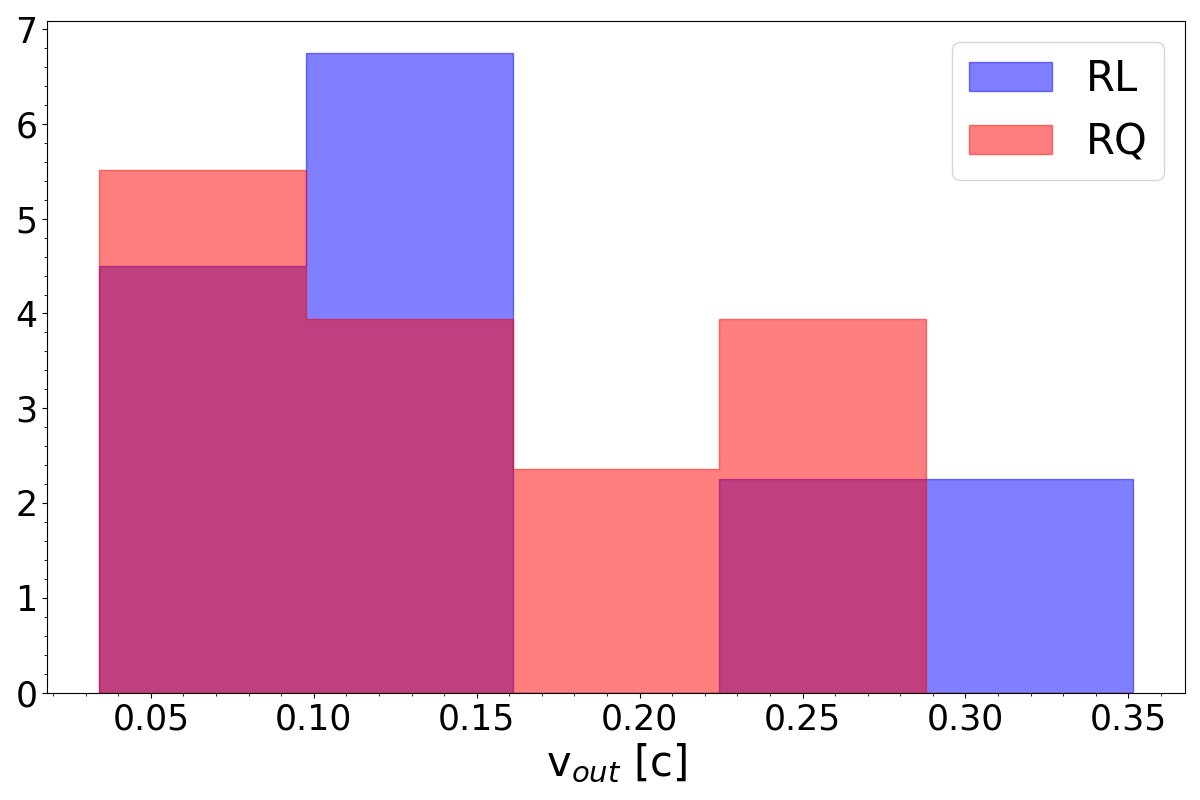}
\includegraphics[width=0.75\columnwidth]{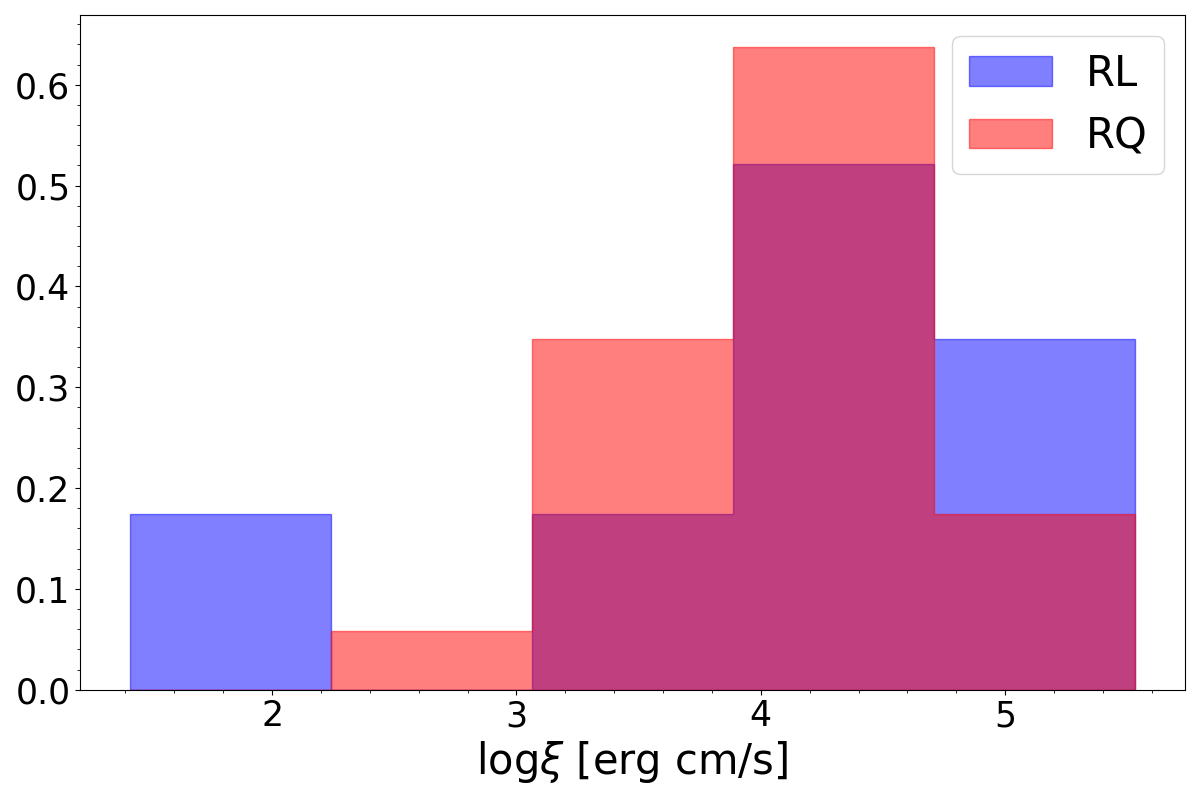}
\includegraphics[width=0.75\columnwidth]{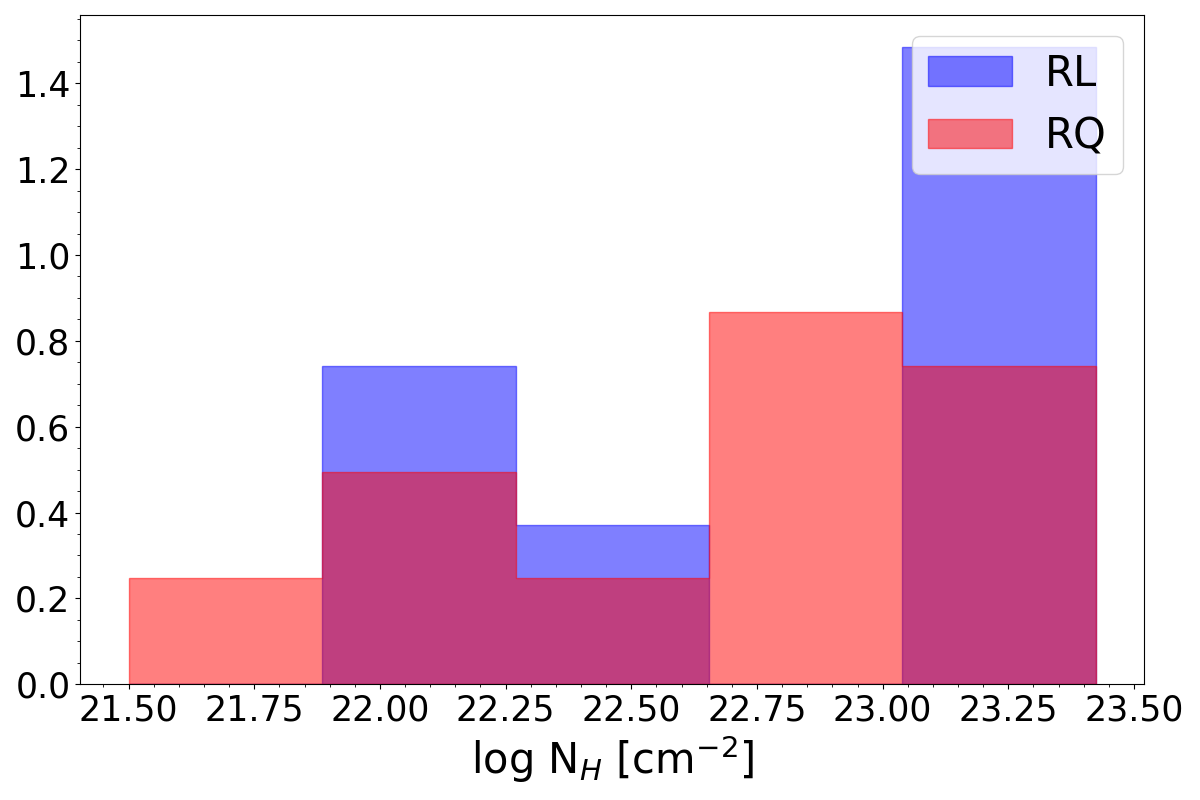}
\caption{Distributions of the logarithm of the velocity in units of the speed of light (upper panel), ionization parameter (central panel), and column density (lower panel) for the outflows detected in the combined sample of RL (blue) and RQ (red) sources.}
\label{distrib2}
\end{figure}

\begin{figure}
\centering
\includegraphics[width=0.75\columnwidth]{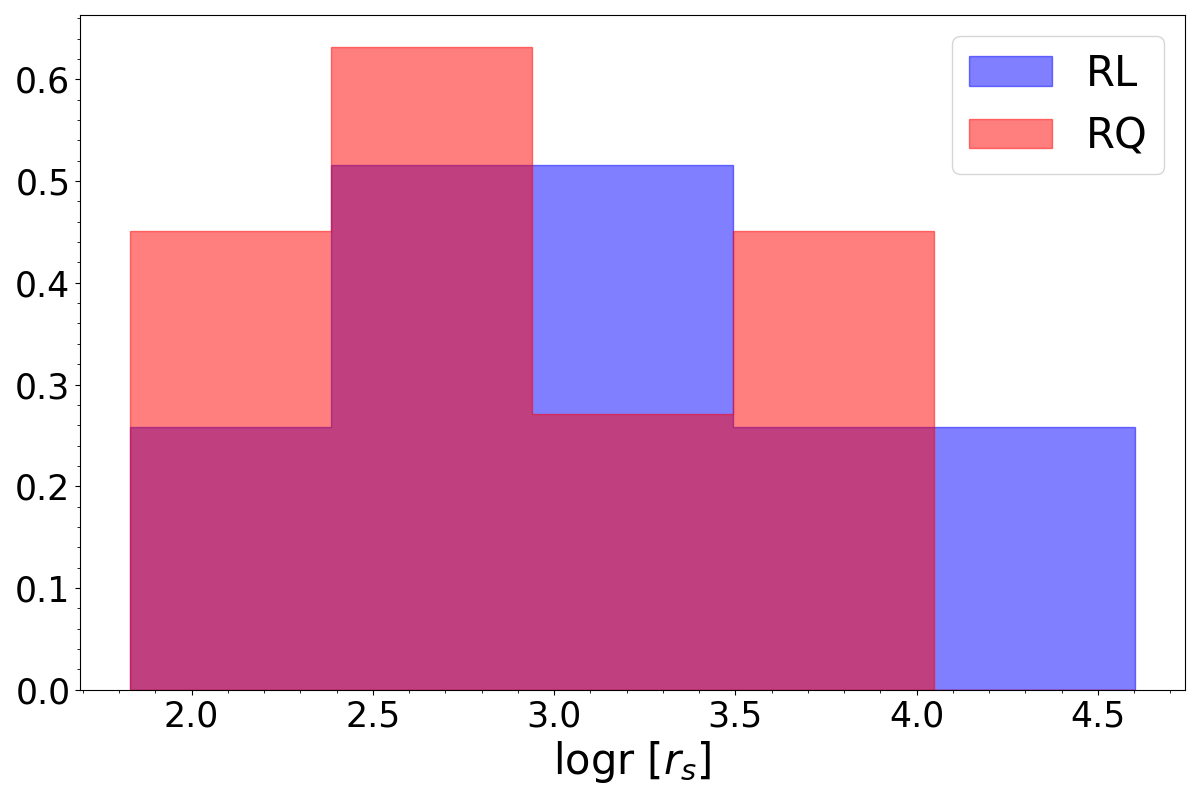}
\includegraphics[width=0.75\columnwidth]{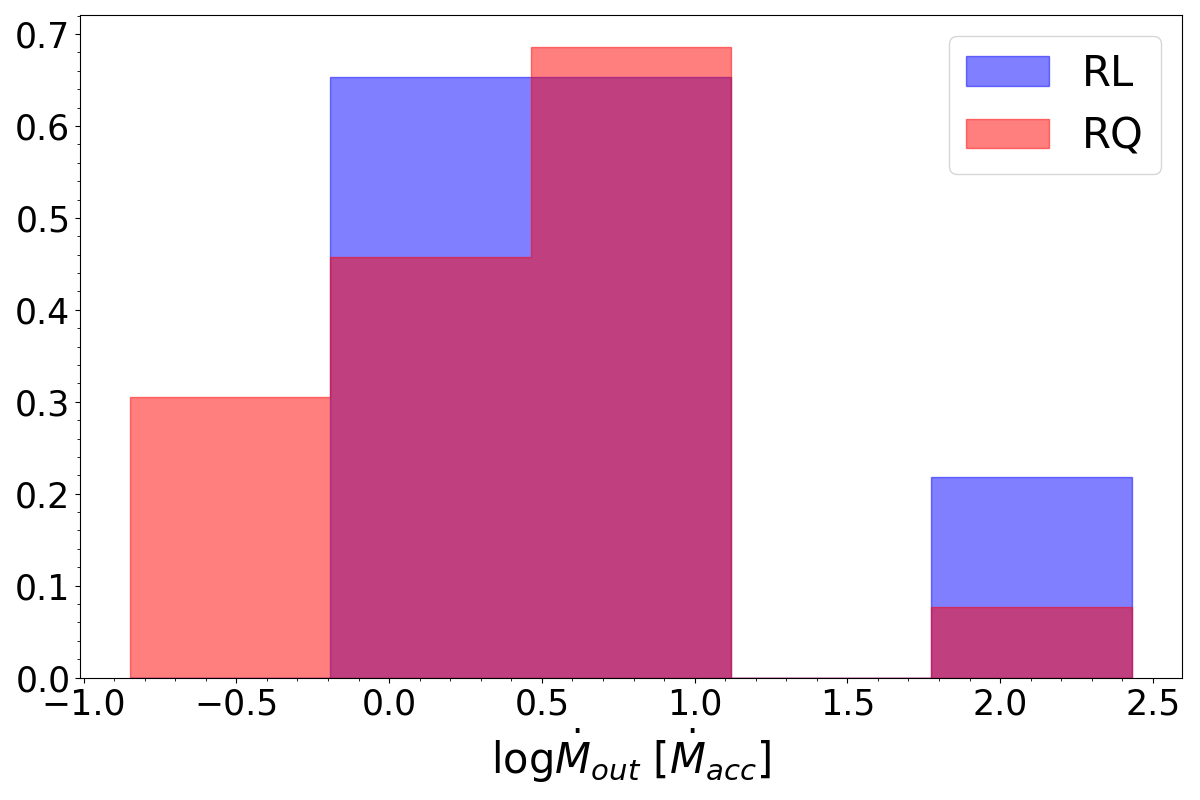}
\includegraphics[width=0.75\columnwidth]{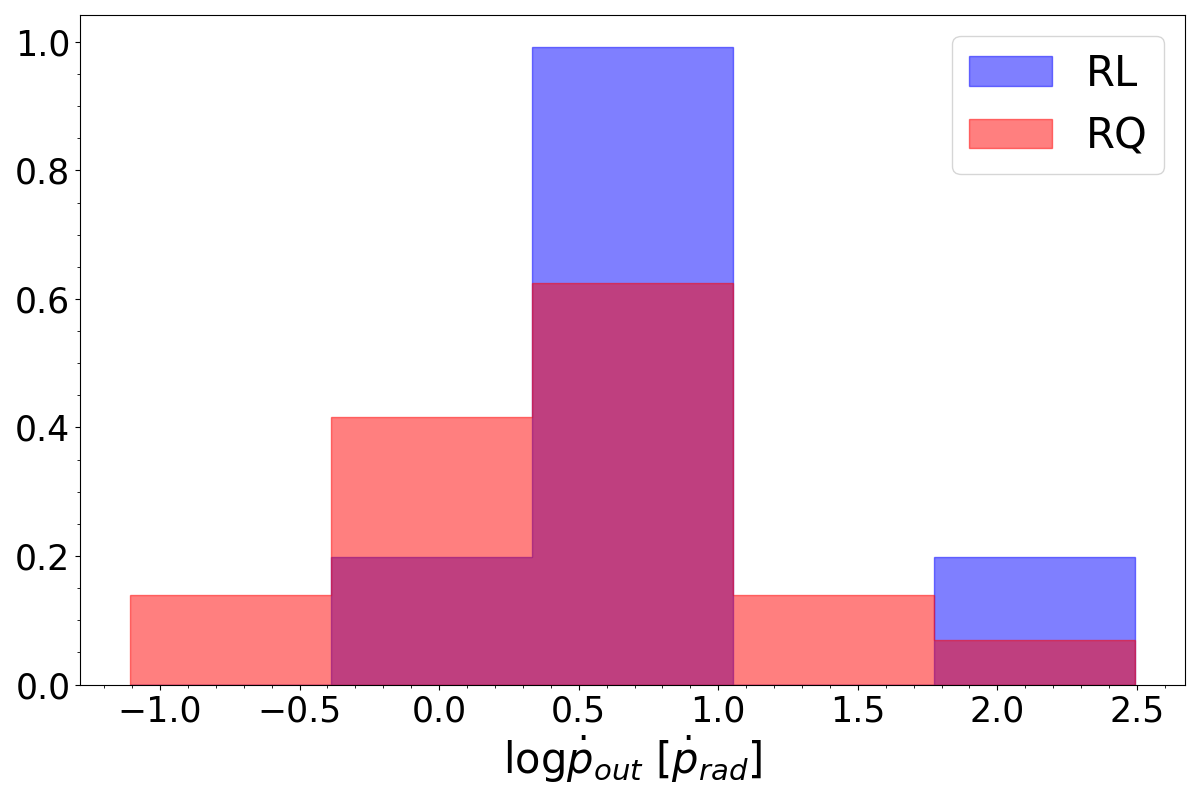}
\caption{Distributions of the logarithm of the wind position in units of the Schwartzschild radius (upper panel), logarithm of the ratio of the mass outflow rate normalized to the Eddington accretion rate (central panel), and the ratio of the wind momentum rate normalized to the momentum rate of the radiation (lower panel) for the combined sample of RL (blue) and RQ (red) sources with detected outflows.}
\label{distrib3}
\end{figure}

The distributions of the UFO parameters, i.e.,  $v_{out}$, log$\xi$ and $N_{H}$, are shown in Fig \ref{distrib2} while the normalized $r_{out}$, $\dot{M}_{out}$ and $\dot{p}_{out}$ are shown in Fig. \ref{distrib3}. A clear superposition is observed in the parameter space of the two classes, and indeed the KS-test can not reject the null hypothesis. Of particular interest is the pronounced correlation between the $v_{out}$ distributions, where the confidence level in rejecting the null hypothesis is only 1\%. These results strongly suggest a common underlying origin for the disc winds, most likely caused by the same physical mechanism(s) in both X-ray bright RQ and RL AGNs. In particular, if we did not know the radio jet properties of these sources, the two classes would be virtually indistinguishable from the point of view of accretion and wind properties alone. This indicates that the accretion disc properties of luminous RQ and RL AGN are rather similar, being both radiative efficient and capable of producing powerful winds, as expected in a self-similar CCA scenario (the multiphase feeding rain does not distinguish between jet or non-jetted feedback; \citealt{Gaspari17}). The dichotomy in the radio jet properties, instead, may be dominated by other parameters, such as the SMBH spin parameter.\\
It is interesting to note that the similarities, and lack of a clear dichotomy, between RQ and RL AGN in our study is supported by the fact that BLRGs in our sample are optically classified as high-excitation galaxies (HEG). This sub-class of RL AGN is most likely associated with the presence of cold accreting material, similarly to the Seyfert case for RQ AGN. Instead, radio-galaxies classified as low-excitation galaxies (LEG) would be powered by hot gas, and typically exhibit lower Eddington ratios than HEG. The high temperature of the accreting gas in LEG would account for the lack of "cold" structures, i.e. molecular torus and broad line region, for the reduced radiative output of the accretion disk, and for the lower gas excitation (e.g., \citet{Best05}; \citet{Buttiglione2010}). This distinction is then more based on the feeding of the SMBH, instead of the morphological classification based on the large-scale radio jet and its feedback.\\

\begin{figure}
\includegraphics[width=\columnwidth]{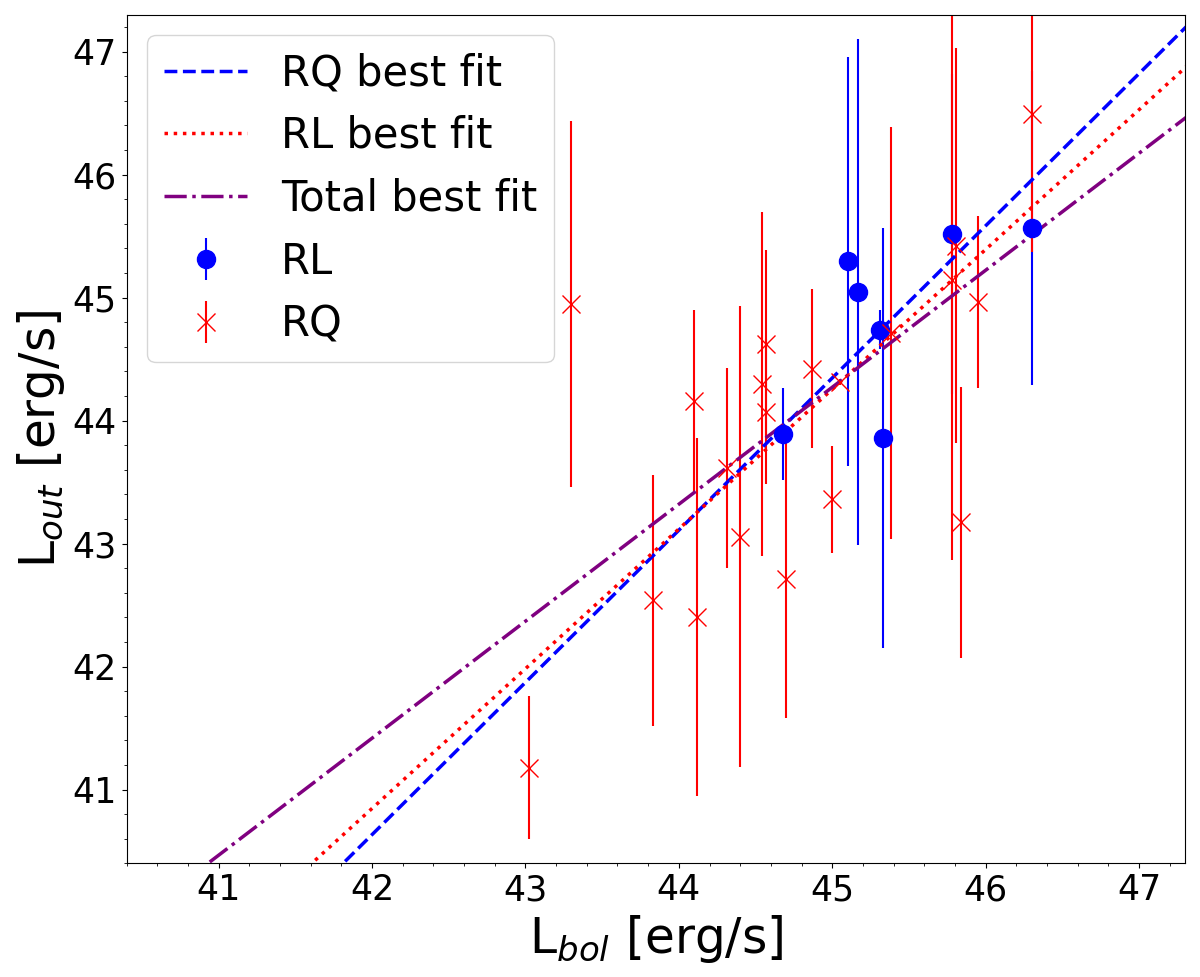}
\caption{Scatter plot showing the disc winds kinetic power versus the bolometric luminosity for the sample of RQ (red cross) and RL (blue circle) AGNs. Using the same color-coding the best fit obtained using a linear regression algorithm are also shown as a dashed blue line and dotted red line for RL and RQ AGNs respectively.}
\label{LL}
\end{figure}

Fig.~\ref{LL} shows the correlation between the wind kinetic power and the bolometric luminosity for both RQ and RL AGNs. We adopt a hierarchical Bayesian model for linear regression, \texttt{linmix} (\citealt{Kelly07, Gaspari19}), to fit the data, taking into account the relative confidence range for each point. The form of the regression is  $log(y)= \alpha + \beta log(x) + \epsilon$ , where $\alpha$, $\beta$ and $\epsilon$ are the intercept and slope coefficients and intrinsic random scatter of the regression, respectively. In this approach, the intrinsic scatter is treated as a free parameter and $\epsilon$ is assumed to be normally distributed with mean zero and variance $\sigma$. The regression model is then fit using Markow Chain Montecarlo (MCMC) sampling where the estimated parameters are obtained as a mean value of the chain. The algorithm also yields an estimate of the intrinsic correlation between the variables, "x-y corr", which comes from the knowledge of the posterior distribution of each physical parameter. This output has values in the interval [-1,1] where positive values point at a positive correlation and viceversa. The derived values are reported in Tab. \ref{DAT}. We observe an equivalent linear correlation in log-log space between L$_{out}$ and L$_{out}$ for both AGN types, which is to say that the kinetic power of the outflows is directly proportional to the luminosity for both classes. The two linear regressions models are consistent, however it is not possible to determine a significant relation with the constraints on the RL AGNs sample compared to those of the RQ, due to the current limited number of objects.

\begin{table}
\caption{Correlation analysis of outflow luminosity with respect to the AGN bolometric luminosity for RL, RQ and entire sample. The first two column represent the regression coefficients whereas the third the regression intrinsic scatter for the two AGN populations.}            
\label{DAT}      
\centering                          
\begin{tabular}{l c c c c}        
\hline\hline                 
Type & $\beta$  & $\alpha$ & $\sigma$ & x-y corr \\    
\hline                       
RL & 1.07$\pm$0.76 & -3.87$\pm$8.16 & 1.29$\pm$1.34 & 0.49$\pm$0.48\\
RQ & 1.03$\pm$0.30 & -2.19$\pm$3.26 & 0.56$\pm$0.29 & 0.85$\pm$0.16\\
\hline           
\end{tabular}
\end{table}

Our analysis up to this point suggests that accretion disc winds are not only a common trait of both types of AGNs, but are also most likely produced by the same physical mechanisms and conditions. Consequently, the radio-loudness dichotomy seems not to be a good tracer for the presence or lack of outflows, nor does it contain information on the main parameters of the wind. 

\subsection{Disc wind characteristics in the entire AGN sample} \label{char}

In the previous section, we deduced that the overall physical parameters of the outflows are consistent between RQ and RL AGNs. Therefore, hereafter, we will consider them both as a single population. In Tab. \ref{table:2} the mean parameters for the UFOs in the entire sample are shown. 

\begin{table}
\caption{Mean values of the AGNs intrinsic properties and disc wind parameters for the entire sample.}         
\label{table:2}      
\centering                          
\begin{tabular}{l r r}        
\hline\hline                 
Parameter  & value  \\    
\hline    
log($M_{BH}$) & $7.8\pm0.8$\\ 
log($L_{bol}$) & $44.8\pm0.9$\\
log$\lambda_{Edd}$ & $-1.0\pm0.5$\\
\hline
log$\xi$ & $0.6\pm0.1$ \\
log$N_H$ & $22.9\pm0.7$ \\
$v_{out}/c$ & $0.15\pm0.08$ \\
log($r_{out}/r_{s}$) & $3.0\pm0.7$ \\  
log($\dot{M}_{out}/\dot{M}_{acc}$) & $0.4\pm0.7$ \\
log($\dot{M}_{out}/\dot{M}_{Edd}$) & $-0.6\pm0.5$ \\
log($L_{out}/L_{Edd}$) & $-1.7\pm0.7$\\ 
log($\dot{p}_{out}/\dot{p}_{rad}$) & $0.5\pm0.7$\\ 

\hline                                  
\end{tabular}
\end{table}

The distributions of log$N_H$ and log$\xi$ span ranges of $21<$log($N_H$)$<24$ and $1<$log($\xi$)$< 6$, respectively. Their mean values are reported in Tab. \ref{table:2} and are consistent with the idea that disc winds are characterized by a substantial column density and the gas is highly ionized. Moreover, it can also be observed that the average outflow velocity, $v_{out}$, is mildly relativistic, with an average value of $\simeq$15\% of the speed of light.\\ The Eddington ratio can be considered a useful proxy for the state of an AGN accretion disc. The bolometric luminosity in our sample is lower than the Eddington luminosity, with a mean value log$\lambda_{Edd}=-1.0\pm0.5$, suggesting that the accretion and radiation emission, which may be a possible mechanism to accelerate winds, have found an equilibrium. While the possibility of a highly luminous and transient phase, which would eventually sweep away all the matter in the proximity of the BH setting forth the end of the AGN phase, cannot be ruled out, it should also be considered that the Eddington limit is an approximation for spherical emission and, as such, could underestimate the critical value of a beamed emission.\\

The distribution of the radial distance from the BH expressed in physical units ($cm$) spans six orders of magnitude, due to the wide interval of observed $M_{BH}$ and $v_{out}$. However, this distribution becomes narrower when normalizing it to the respective $r_s$, with a mean value of log($r_{out}/r_{s}$) $=3.0\pm0.7$, i.e. approximately $\sim0.0003-0.03$ pc. This suggests that Fe K absorbers are swirling closer to the BH than the traditional soft X-ray warm absorbers, which are frequently observed at pc scales and beyond (e.g \citet{Crenshaw03}, \citet{Blustin05} and \citet{Kaastra12}).\\ Our results agree with those found in the literature, in which the general consensus is that the launch of highly ionized outflows is triggered in the inner regions of accretion discs (e.g., \citet{Proga04}, \citet{Schurch09}, \citet{Sim08} and \citet{King10}).\\ 

The mean mass outflow rate ranges from 0.01 to 1 $M_{\odot}$/yr. However, when normalizing it to $\dot{M}_{Edd}$, the interval is $\sim0.1-1 \dot{M}_{Edd}$.  Most of the sources in our sample present a $\dot{M}_{out}<\dot{M}_{Edd}$, demonstrating the consistency of using the Eddington limit as an upper limit or, at the very least, as an extreme value to which compare our results.\\ The observed disc winds are able to sweep and transport large amounts of gas, corresponding to an average mass outflow rate of $\sim$25\% the Eddington accretion rate. The mass outflow rate distribution is also examined as a function of the mass accretion rate, $\dot{M}_{acc}$, where its average value is $0.4\pm0.7$, as shown in Tab. \ref{table:2}. The data is mostly consistent with the unitary value, so that $\dot{M}_{out}$ and $\dot{M}_{acc}$ have the same order of magnitude. This means that what flows in at the micro scale is roughly comparable to what is re-ejected back in outflows\footnote{We note that $\dot{M}_{acc}$ is the micro feeding rate linked to the inner disc, not the mass rate through the BH horizon.}. In other words, the feeding and feedback are tightly and efficiently self-regulated. This accretion and ejection cycle is a fundamental element in AGN research, as also evidenced by the growing body of literature (\citet{Fiore17,Gaspari20} and references therein) that describes how this process most likely controls the SMBH - host galaxy system. This interaction may also be one of the explanations to the relation between $M_{BH}$ and the velocity dispersion of the galaxy's bulge (e.g see \citet{Pounds14}). Therefore, our results are consistent with the present theory of a connection between the activity of the AGN and the host-galaxy medium. In particular, our results are well consistent with the predictions of a CCA self-regulated duty cycle, which has been shown to drive ultrafast outflows of the order of $\sim$\,$0.1 c$ velocities and mass outflow rates comparable to the disc inflow rates \citep{Gaspari17}.\\

The kinetic power of the outflows is quite high, with values between $\simeq 10^{42}$ and $10^{46}$ erg~s$^{-1}$, and a normalized mean value of log($L_{out}/L_{Edd}$) $= -1.7\pm0.7$. The latter corresponds to $\simeq$3\% of the Eddington luminosity. According to models of AGN feedback, a wind power of $\sim$0.1-1$\%L_{Edd}$ must be converted into mechanical power in order to drive significant effects on the co-evolution of both SMBH and the host galaxy (see \citet{DiMatteo05}; \citet{Hopkins10}). Our results show the the most energetic disc winds in both RQ and RL AGNs, may indeed have an high impact on AGNs feedback.

The average value of the normalized momentum rate (or force) of the outflows, i.e. $\dot{p}_{out}\sim 3~\dot{p}_{rad}$ is provided in Tab. \ref{table:2}. The two physical parameters have the same order of magnitude, suggesting that radiation pressure may be an essential component in the acceleration of these winds. One candidate for the acceleration of such highly ionized outflows is Compton scattering of the UV and X-ray continua. However, due to the relatively small cross section of highly ionized gas, this process is not very efficient and requires both a high luminosity and column density. Moreover, the existence of outflows with greater ratios points at the possibility that another acceleration mechanism besides radiation-pressure may be present, with MHD effects as it potential origin (\citet{Fukumura10}).

In light of our results, given that \(\dot{p}_{out} = \dot{M}_{out} v_{out}\) and \(\dot{p}_{rad} = L_{bol}/c =  \eta c\dot{M}_{acc}\), then \(\beta = v_{out}/c \sim \eta\). Therefore, a positive correlation seems to hold for the velocity of the outflow and the mass-energy conversion efficiency, where the latter is a spin and accretion rate dependent parameter (\citet{Thorne74}). This relation is apparently consistent with the observations, as the mean velocity observed for UFOs is $v_{out}\sim$0.1$c$ \citet{Tombesi10} while $\eta$ is usually estimated to be $\sim$0.1. However, the magnetic driving mechanism, suggested to partially explain the acceleration of disc winds, is tied to the BH spin, while the latter was assumed null in the previous $\eta$ assumption, leading to an inconsistency. In \citet{Madau04} the radiative efficiency as a function of the BH normalized accretion rate and spin is obtained from numerical integration of the relativistic slim disc equations, showing that high values of $\eta$ are attained for highly spinning BH with low $\dot{M}_{acc}$. Still, to further constrain the correlation between $v_{out}$ and $\eta$, high quality data and progress in the models adopted to parametrize both UFOs and the conversion efficiency in AGNs are required. 

\subsection{Correlations with bolometric luminosity} \label{Lbol}

In order to further understand the physics behind our results, the correlation between the outflow physical and scale-invariant properties and the AGN bolometric luminosity was also analyzed. A similar attempt was previously performed by \citet{Gofford15}. We use the same linear regression model adopted in Sec.~\ref{origin} to fit the data. Our results are gathered in Tab. \ref{table:3}.\\

\begin{table*}
\caption{Summary of the correlation analysis of the outflow parameters with respect to the AGN bolometric luminosity. The output in the table are computed using a hierarchical Bayesian model for linear regression as explained in Sec.~\ref{origin}.}         
\label{table:3}      
\centering                          
\begin{tabular}{l c c c c c}        
\hline\hline                 
Parameter & $\beta$  & $\alpha$ & $\sigma$ & x-y corr\\    
\hline    
log($M_{BH}$) & 0.82$\pm$0.12 & -29.36$\pm$4.78 & 0.52$\pm$0.08 & 0.83$\pm$0.08\\ 
log$\lambda_{Edd}$ & 0.17$\pm$0.12 & -8-77$\pm$5.15 & 0.52$\pm$0.08 & 0.39$\pm$0.21\\
\hline
log$\xi$ & -0.01$\pm$0.03 & 0.89$\pm$1.28 & 0.12$\pm$0.01 & -0.05$\pm$0.23\\
log$N_H$ & -0.08$\pm$0.14 & 26.53$\pm$6.29 & 0.64$\pm$0.10 & -0.13$\pm$0.22\\
log($v_{out}/c$) & 0.06$\pm$0.07 & -3.87$\pm$3.39 & 0.31$\pm$0.05 & 0.19$\pm$0.22\\
log($L_{out}$) & 1.11$\pm$0.23 & -5.59$\pm$10.53 & 0.36$\pm$0.21 & 0.92$\pm$0.09\\
log($L_{out}/L_{Edd}$) & 0.20$\pm$0.20 & -10.77$\pm$10.95 & 0.43$\pm$0.20 & 0.34$\pm$0.41\\ 
log($r_{out}$) & 0.57$\pm$0.22 & -9.41$\pm$10.08 & 0.30$\pm$0.17 & 0.84$\pm$0.17\\
log($r_{out}/r_{s}$) & -0.25$\pm$0.26 & 14.14$\pm$11.69 & 0.50$\pm$0.19 & -0.40$\pm$0.38\\ 
log($\dot{M}_{out}$) & 0.83$\pm$0.22 & -37.68$\pm$9.70 & 0.28$\pm$0.16 & 0.92$\pm$0.10\\
log($\dot{M}_{out}/\dot{M}_{acc}$) & -0.17$\pm$0.22 & 8.17$\pm$9.84 & 0.28$\pm$0.16 & -0.40$\pm$0.49\\
log($\dot{p}_{out}$) & 0.98$\pm$0.22 & -9.16$\pm$9.80 & 0.28$\pm$0.17& 0.94$\pm$0.08\\
log($\dot{p}_{out}/\dot{p}_{rad}$) & -0.01$\pm$0.22 & -0.05$\pm$0.22 & 0.27$\pm$0.16 & -0.02$\pm$0.58\\

\hline           
\end{tabular}
\end{table*}

We first point out that the regression slope is generally considered a good proxy of the correlation between independent, in our study $L_{bol}$, and dependent  variables. However, in our regression algorithm the linear correlation coefficient, x-y corr, is directly estimated through MCMC sampling. This parameter yields the true correlation between the variables and is more reliable than the regression slope. In fact, in complex distributions a weak linear fit, given by a positive $\beta$, may correspond to no correlation, given by a x-y corr consistent with 0. While we adopted a linear regression model to fit our data, some distributions may differ from a simple linear trend. This information is partially contained in the $\sigma$ coefficient which quantifies the intrinsic scatter of the data with respect to the linear regression model. As different $\sigma$ values may affect the uncertainties of the best fit parameters, we also investigate the possible scaling relations between the two. We found no statistical evidence of a correlation between $\sigma$ and the uncertainties, i.e. an higher $\sigma$ does not lead to an higher uncertainty in the best fit parameters.\\

Although the intrinsic scatter on $M_{BH}$ is relatively high, i.e. 0.52$\pm$0.08, a positive correlation with $L_{bol}$ can be clearly observed, with x-y corr$=0.83\pm0.08$, highlighting the fact that more massive AGNs possess greater luminosities, hence stronger feeding/CCA rates. A strong positive correlation is found between $L_{bol}$ and $\lambda_{Edd}$, as x-y corr = $0.92\pm0.09$. This trend suggests that more luminous AGNs are more efficient in accelerating UFOs.\\ The correlation coefficients of $N_H$ and $\xi$ are consistent with zero, pointing at both parameters being independent from $L_{bol}$. The intrinsic scatter in both fit is $\geq$0.6, which implies that linear regression is not optimal to model the data distribution.

Of particular interest is that $v_{out}$ and $L_{bol}$ seem to be weakly dependent, with a x-y corr$=0.19\pm0.22$. As radiation pressure is considered to play a relevant role in the acceleration of the outflows, one may expect to observe a positive correlation as in \citet{Gofford15} where $\beta=0.4^{+0.3}_{-0.2}$. Nonetheless, it should be noted that in this work the selection criteria used for the sample is different from the one used in the previously cited work. In particular, we are considering only outflows with velocities greater than 1\% of the speed of light to focus the analysis on the most powerful accretion disc (i.e. UFOs) components. Moreover, a possible explanation for the lack of or weakness of such relation is that the instantaneous outflow velocity is simply not a good proxy of its acceleration, as it does not account for the mass flux. Instead, parameters that better represent the wind acceleration are the mass outflow rate and momentum rate, as explained below. 

\begin{figure}
\centering
\includegraphics[width=0.8\columnwidth]{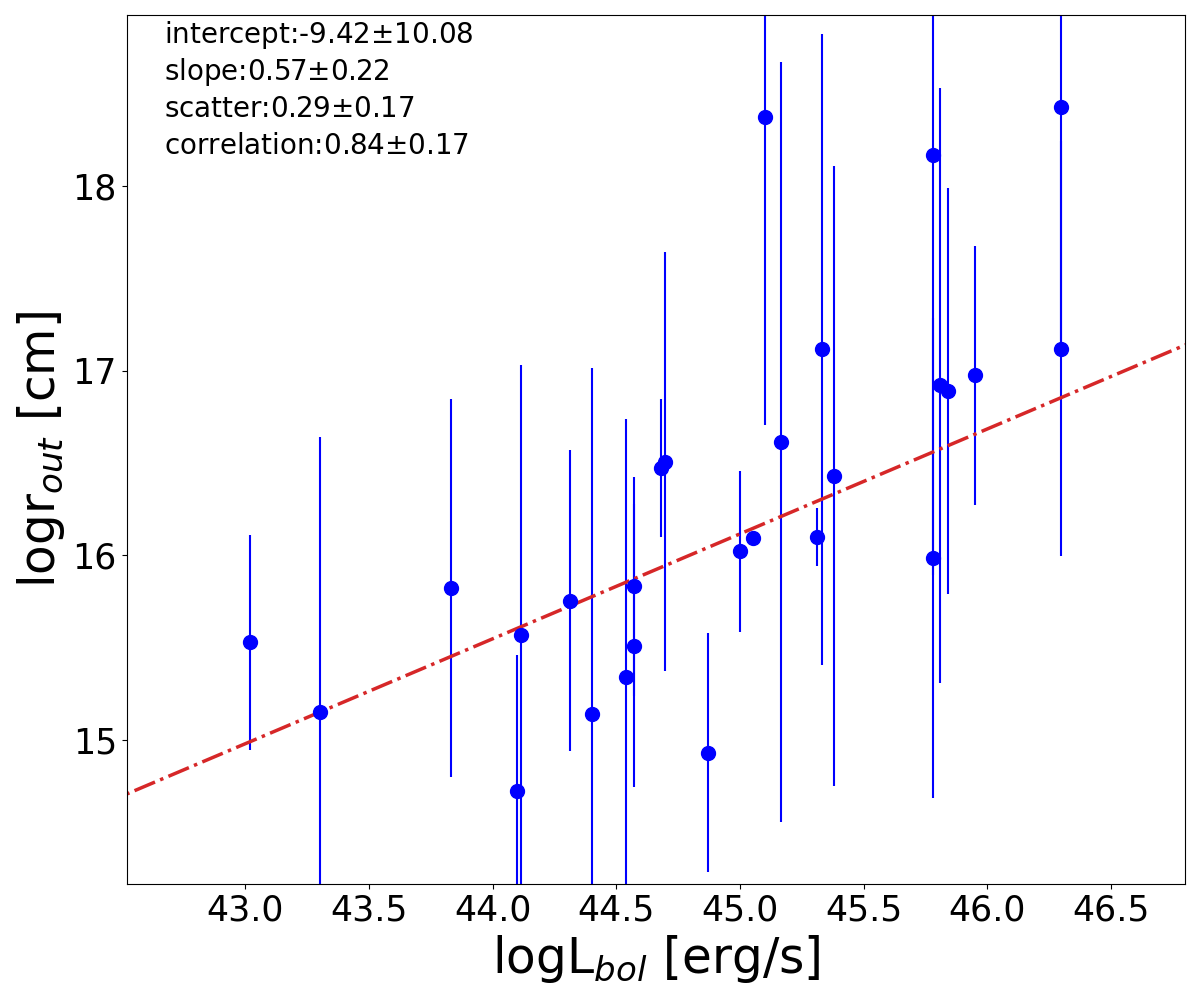}
\includegraphics[width=0.8\columnwidth]{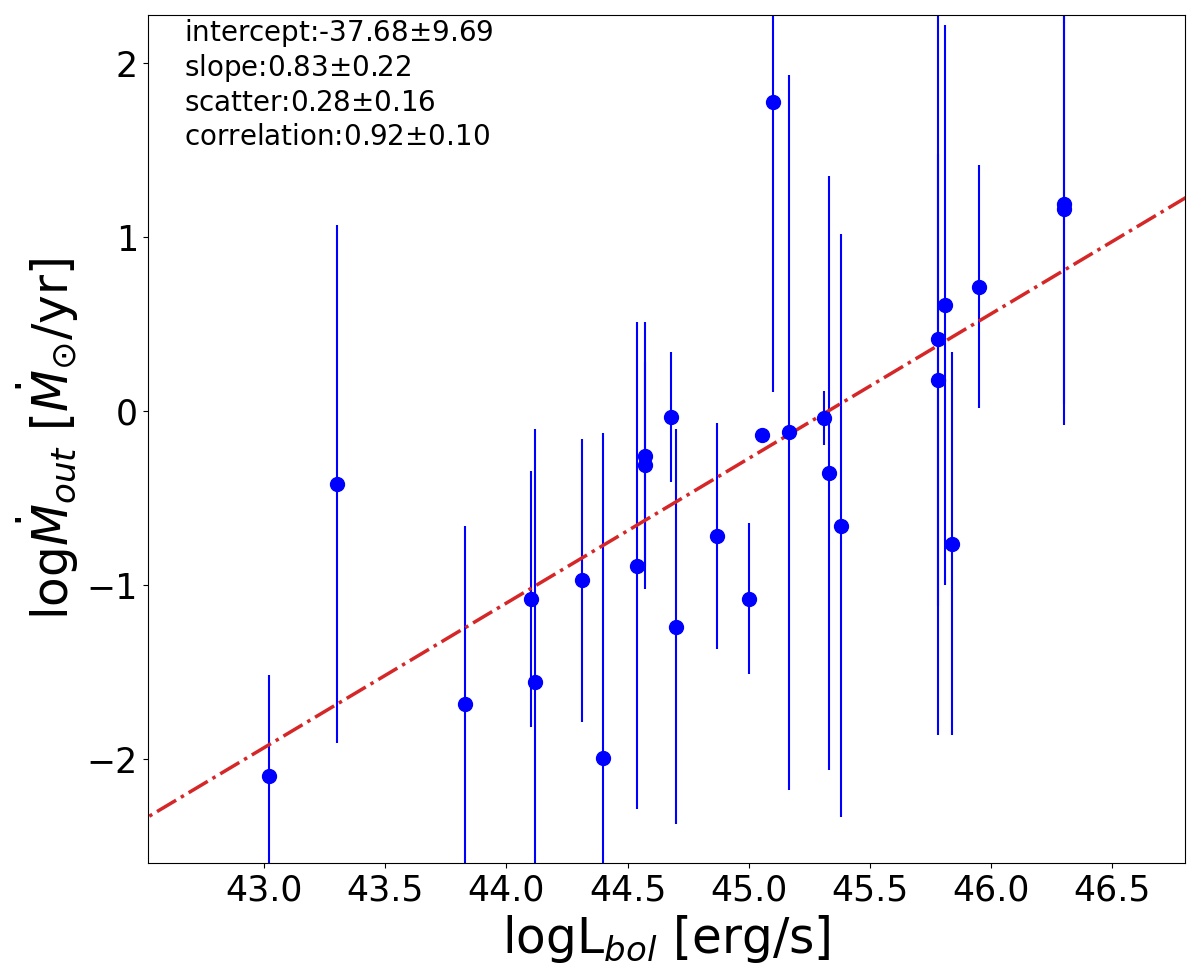}
\includegraphics[width=0.8\columnwidth]{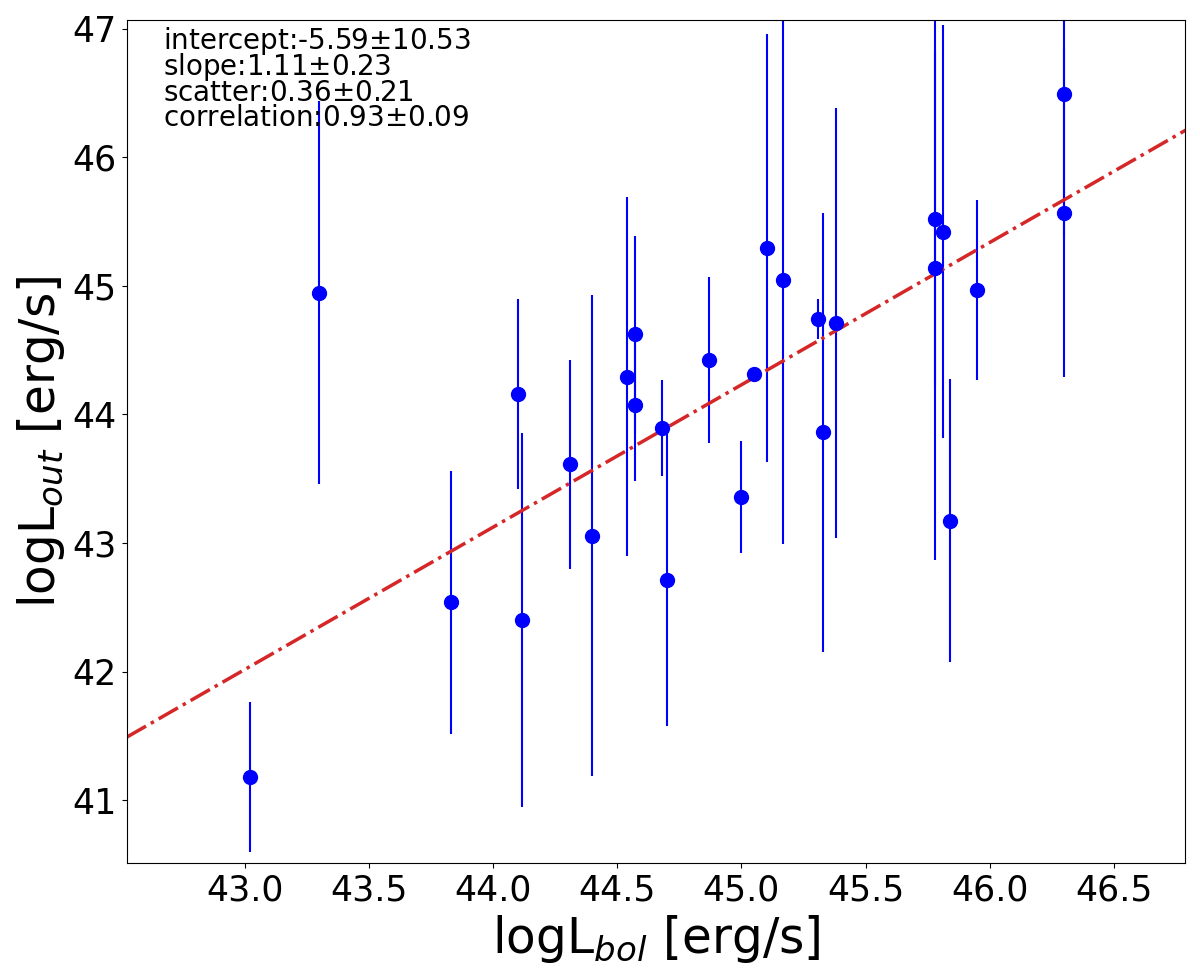}
\caption{Linear regression fit for the outflows physical parameters as a function of $L_{bol}$. In figure the outflow radial distance (upper), mass outflow rate (central) and instantaneous wind power (lower) are shown in physical units in a logarithmic plane. The best fit linear regression for the data is also highlighted in each panel as a dash-dotted red line.}
\label{compL1}
\end{figure}

\begin{figure}
\centering
\includegraphics[width=0.8\columnwidth]{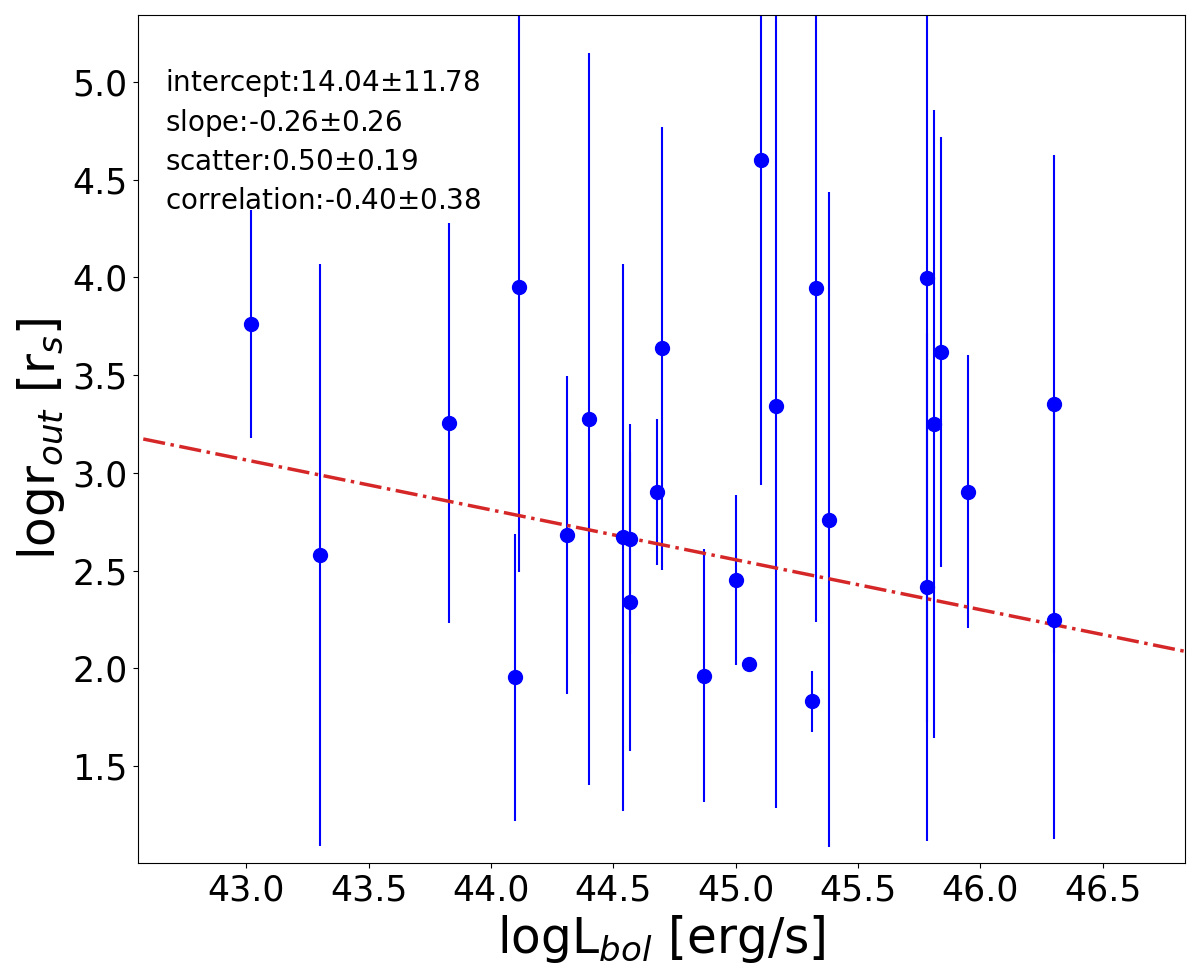}
\includegraphics[width=0.8\columnwidth]{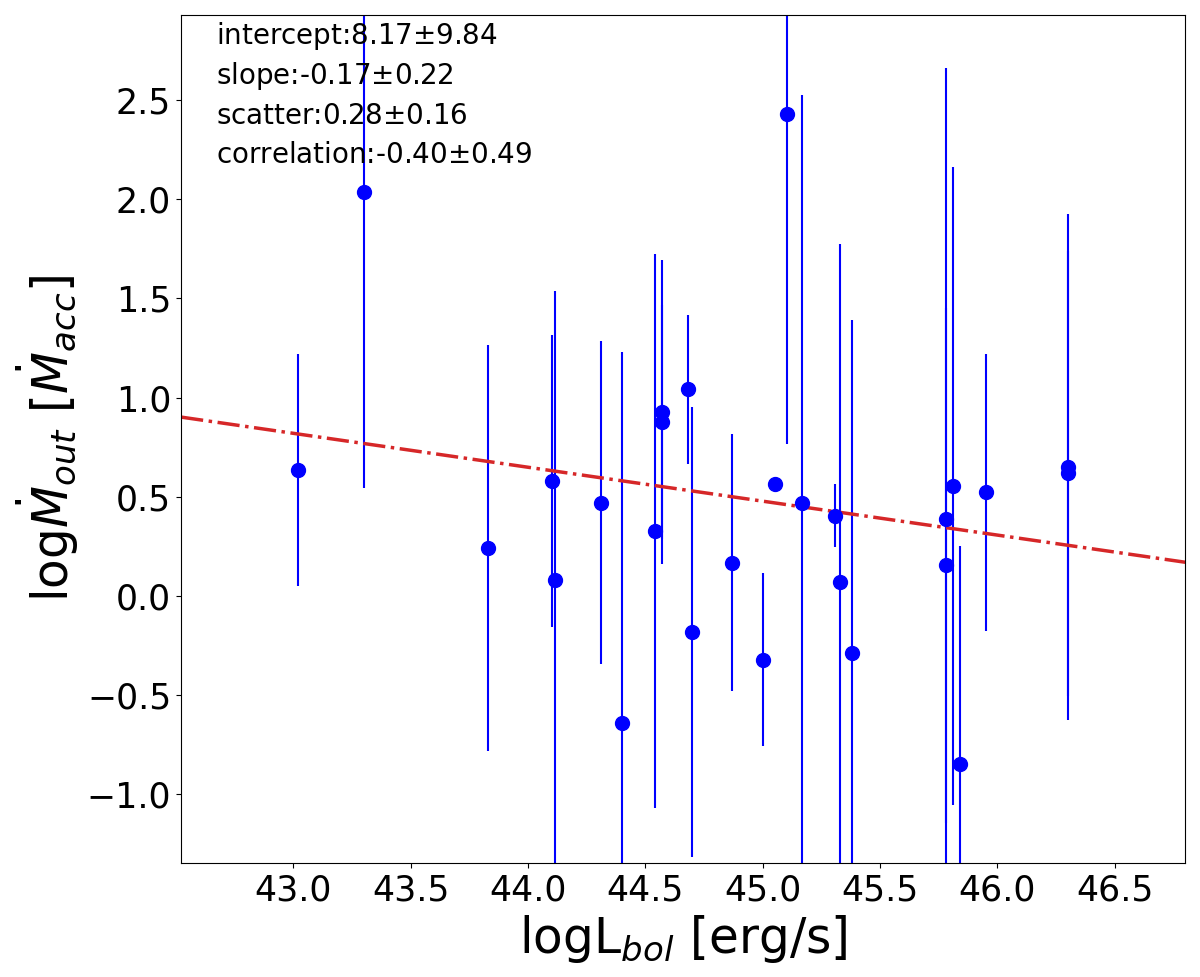}
\includegraphics[width = 0.8\columnwidth]{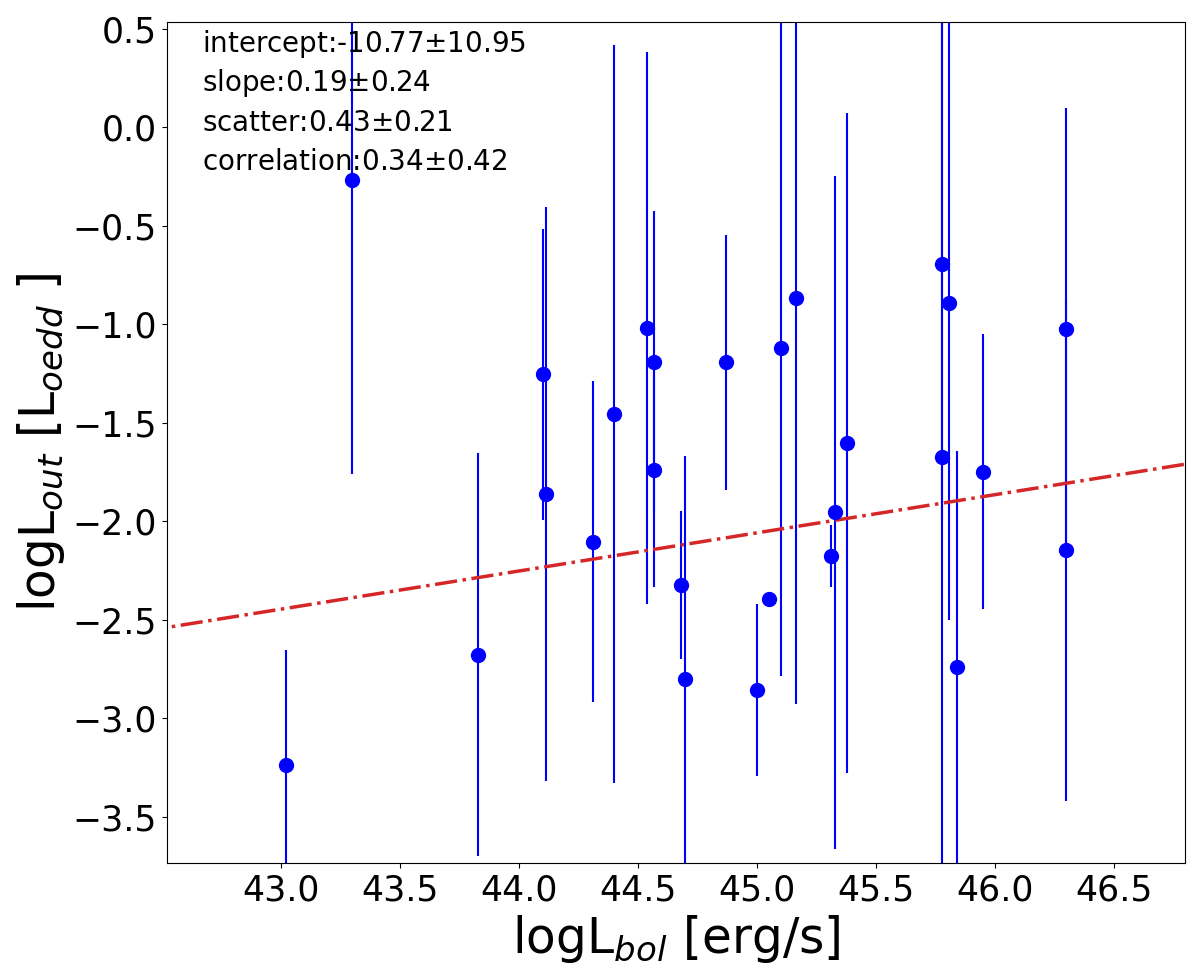}
\caption{Linear regression fit for the outflows normalized parameters as a function of $L_{bol}$. In figure the outflow radial distance (upper), mass outflow rate (central) and instantaneous wind power (lower) are shown in a logarithmic plane. The best fit linear regression for the data is also highlighted in each panel as a dash-dotted red line.}
\label{compL2}
\end{figure}

A statistically significant correlation with $L_{bol}$ (see Fig. \ref{compL1} and Tab. \ref{table:3}) is observed in the outflow radial distance (i.e. $0.84\pm0.17$), mass outflow rate (i.e. $0.92\pm0.10$), momentum rate (i.e. $0.92\pm0.12$), and instantaneous wind power (i.e. $0.94\pm0.08$), respectively. These results are in line with the ones reported in \citet{Gofford15}, \citet{Tombesi12}, and \citet{Tombesi13}. The strong correlation between the parameters which take into account the energetics of the UFOs and the AGNs $L_{bol}$, once again underlines that radiation is a relevant element in the description of the acceleration and evolution of the outflows. In particular, $\dot{p}_{out}$ represents the most relevant tracer of the correlation, with a low intrinsic scatter, $\sigma = 0.28\pm0.16$. The most energetic outflows, and therefore the ones more likely to disrupt the SMBH accretion and interact with the host-galaxies bulges are found in the most luminous AGNs.
Our results are consistent with \citet{Gofford15}, however, the correlation between $L_{out}$ and $L_{bol}$ obtained by the authors is slightly steeper (i.e. $\beta=1.5^{+1.0}_{-0.8}$). Once again this difference may be explained by the different selection criteria used here, as winds with lowest power are not being considered in our study.

An opposite behavior is observed in the normalized outflows parameters (see Fig. \ref{compL2} and Tab. \ref{table:3}). These scale invariant properties are either not or weakly anti-correlated to $L_{bol}$. Specifically, the correlation coefficient of the normalized momentum rate $\dot{p}_{out}/\dot{p}_{rad}$ is consistent with 0, while the normalized  outflow distance $r_{out}/r_s$ and mass outflow rate $\dot{M}_{out}/\dot{M}_{acc}$ and anti-correlated. The intrinsic scatter for each value are higher or similar to the ones obtained in the non rescaled analysis. These evidences suggest that outflows are not related to the bolometric luminosity of the AGNs once the dependence on the scale of the system, i.e. the BH mass, is removed.\\ These findings are still consistent with the analysis from \citet{Gofford15}, however, an anti-correlated trend is observed in most of our parameters. Our results support the general idea that UFOs may take on various scales and energetic primarily determined by the intrinsic sizes of their AGN hosts, but are produced by the same underlying mechanism. These picture can be better understood in the CCA framework (Sec.~\ref{intro}), in which multiphase clouds condense out of the diffuse macro-scale halos of galaxies/groups of galaxies and trigger self-regulated AGN outflow feedback, as found here. Such raining clouds, not only drive the observed intrinsic clumpiness of the meso AGN (Sec.~\ref{data}), but also allow for significant disc accretion rates comparable to the outflow rates (Sec.~\ref{char} and Table \ref{table:2}), which would be unattainable in a hot mode of accretion (e.g., Bondi/ADAF). CCA models predict ultrafast outflows triggered regardless of the radio loudness of the systems, with typical velocities of $0.1 c$ and mechanical efficiencies of a few percent \citep{Gaspari17}, consistently with our findings (Sec.~\ref{origin}). Such combined evidences suggest that self-similar CCA is a key mechanism shaping the evolution of our observed systems.

\subsection{Correlations with Eddington Ratio and Radio Loudness}

\begin{table*}
\caption{Summary of the correlation analysis of the outflow parameters with respect to the Eddington ratio and the radio loudness parameter. The output in the table are computed using a hierarchical Bayesian model for linear regression as explained in Sec.~\ref{origin}.}         
\label{table:4}      
\centering                          
\begin{tabular}{l c c c c c}        
\hline\hline                 
Parameter & $\beta$  & $\alpha$ & $\sigma$ & x-y corr\\    
\hline    
& & $\lambda_{Edd}$ correlations & &\\
\hline
log($L_{out}$) & -0.13$\pm$0.60 & 43.92$\pm$0.73 & 0.96$\pm$0.27 & -0.07$\pm$0.32\\
log($L_{out}/L_{Edd}$) & 0.51$\pm$0.41 & -1.42$\pm$0.55 & 0.37$\pm$0.20 & 0.55$\pm$0.38\\ 
log($r_{out}$) & 0.04$\pm$0.39 & 16.13$\pm$0.53 & 0.33$\pm$0.20 & 0.04$\pm$0.53\\
log($r_{out}/r_{s}$) & 0.74$\pm$0.43 & 3.50$\pm$0.57 & 0.45$\pm$0.19 & 0.61$\pm$0.31\\ 
log($\dot{M}_{out}$) & -0.26$\pm$0.45 & -0.74$\pm$0.58 & 0.54$\pm$0.24 & -0.26$\pm$0.41\\
log($\dot{M}_{out}/\dot{M}_{Edd}$) & 0.46$\pm$0.37 & -0.25$\pm$0.51 & 0.29$\pm$0.16 & 0.57$\pm$0.40\\
log($\dot{M}_{out}/\dot{M}_{acc}$) & -0.56$\pm$0.38 & -0.27$\pm$0.53 & 0.29$\pm$0.16 & -0.64$\pm$0.33\\
log($\dot{p}_{out}$) & -0.18$\pm$0.51 & 34.65$\pm$0.65 & 0.72$\pm$0.25& -0.14$\pm$0.36\\
log($\dot{p}_{out}/\dot{p}_{rad}$) & -0.51$\pm$0.37 & -0.19$\pm$0.51 & 0.27$\pm$0.16 & -0.63$\pm$0.37\\
\hline    
& & R$_x$ correlations & & \\
\hline
log($L_{out}$) & 0.51$\pm$0.39 & 45.57$\pm$1.48 & 0.98$\pm$0.37 & 0.43$\pm$0.30\\
log($L_{out}/L_{Edd}$) & 0.04$\pm$0.26 & -2.03$\pm$0.97& 0.57$\pm$0.30 & 0.08$\pm$0.39\\ 
log($r_{out}$) & 0.13$\pm$0.26 & 16.49$\pm$0.98 & 0.55$\pm$0.32 & 0.19$\pm$0.38\\
log($r_{out}/r_{s}$) & -0.34$\pm$0.24 & 1.52$\pm$0.89 & 0.49$\pm$0.28 & -0.55$\pm$0.33\\ 
log($\dot{M}_{out}$) & 0.34$\pm$0.30 & 0.59$\pm$1.11 & 0.69$\pm$0.32 & 0.42$\pm$0.31\\
log($\dot{M}_{out}/\dot{M}_{Edd}$) & -0.14$\pm$0.22 & -1.33$\pm$0.80 & 0.44$\pm$0.25 & -0.32$\pm$0.38\\
log($\dot{M}_{out}/\dot{M}_{acc}$) & -0.03$\pm$0.23 & 0.39$\pm$0.86 & 0.50$\pm$0.25 & -0.07$\pm$0.39\\
log($\dot{p}_{out}$) & 0.43$\pm$0.33 & 36.15$\pm$1.23 & 0.78$\pm$0.34& 0.45$\pm$0.30\\
log($\dot{p}_{out}/\dot{p}_{rad}$) & 0.06$\pm$0.23 & 0.68$\pm$0.85 & 0.49$\pm$0.26 & 0.12$\pm$0.39\\
\hline           
\end{tabular}
\end{table*}

In this subsection we investigate the correlation between wind properties and Eddington ratio and Radio Loudness, using once again a hierarchical Bayesian model for linear regression. We point out that the sources for which we computed $R_x$ are only 17 out of the 27 in our sample, as a radio flux for the source is required. Our results are collected in Table \ref{table:4}.\\ 
Considering the normalized distance from the SMBH, the mass outflow rate and the mechanical power normalized to the Eddington limit, we find a possible proportionality to $\lambda_{Edd}$, with correlation coefficients equal to 0.61$\pm$0.31, 0.57$\pm$0.40 and 0.55$\pm$0.38, respectively. As $\lambda_{Edd}$ is the ratio between $L_{bol}$ and $L_{Edd}$, these trends show once again the relationship between the outflow parameters and the bolometric luminosity discussed in the previous section. Thus, positive correlations are expected as previously discussed. Instead, the mass outflow rate and momentum rate normalize to the radiation pressure are anti-correlated to $\lambda_{Edd}$, with coefficients -0.64$\pm$0.33 and -0.63$\pm$0.37. These findings suggest that the mass driven outwards by outflows decreases as the accretion rate of the SMBH increases. We point out that the values obtained by this analysis may be just lower limits, as this anticorrelation is surely affected by our assumption of constant accretion rate efficiency $\eta$ = 0.1. Although an accurate examination of these processes is beyond the scope of this paper, it may be interesting in the future to compare values of accretion directly obtained through accretion disc fitting models.\\
We now discuss briefly the correlations found between the outflow parameters and the X-ray radio loudness. We observe a weakly positive correlation between the outflow parameters and $R_x$, although characterized by a large intrinsic scatter. Nevertheless, these results may suggest that the radio luminosity, normally attributed to the relativistic jet emission, and the outflow power are connected. The normalized outflows parameter show correlation coefficients that are largely consistent with zero. An exception is the normalized distance $r_{out}$/$r_s$, which seems to be anti-correlated to $R_x$, with correlation coefficient -0.55$\pm$0.33.\\ Understanding the nature of these relationships may be difficult, as our AGN sample is limited in both range of $\lambda_{Edd}$ and $R_x$ values to effectively capture meaningful correlations. More studies with larger samples are required to understand how the efficiency in accretion of the SMBH and the relativistic jet influences the disc and its dynamics.

\section{Conclusions} \label{conc}

In this work, we explored the physical parameters of UFOs through a uniform analysis of a sample of local X-ray bright RQ and RL AGNs. In our statistical analysis we investigated several correlations between different outflow parameters, and with respect to the AGNs bolometric and Eddington luminosities. Our results indicate that accretion disc winds are not only a common trait of both classes of AGNs, but that they are also most likely produced by the same physical mechanism and conditions. Consequently, the radio-loudness dichotomy seems not to be a good tracer for the presence or lack of outflows, nor to be informative of the main parameters of the winds.\\ On average approximately the same amount of material accreted by the SMBH is ejected through disc winds. This evidence is in agreement with the self-regulation of the accretion and ejection in AGN, in particular related to a CCA scenario. The average wind power corresponds to $\simeq$3\% of the Eddington luminosity, indicating that disc winds can indeed provide a significant feedback effect in both RQ and RL AGNs.\\ The outflow parameters related to the energetics are strongly correlated with the bolometric luminosity, highlighting that the most powerful winds are found in the most luminous AGNs, which, most likely, are also highly accreting. Surprisingly, a statistically significant correlation is not found between the outflow velocity and the AGN luminosity, suggesting that velocity alone is not the most relevant wind parameter, but instead we need to consider the total wind power. Moreover, the lack of statistical significant correlations between the normalized outflows parameters and the bolometric luminosity may imply that the underlying UFOs acceleration mechanism(s) are the same for a variety of systems.\\ In the future, it will be interesting to extend a similar study to disc winds detected in both local and high-redshift quasars, to extend the exploration to winds driven by stellar-mass black holes, and to compare the results with detailed numerical simulations. Moreover future missions focused on high-resolution X-ray spectroscopy, like XRISM, will improve exponentially the current statistics of UFO observations in both RL and RQ AGNs, allowing more stringent constraints on the wind parameters. 

\section*{Acknowledgements}
This work is based on results from the Bachelor's thesis in Physics by SM at the Tor Vergata University of Rome. FT acknowledges funding from the European Union - Next Generation EU, PRIN/MUR 2022 (2022K9N5B4). MG acknowledges support from the ERC Consolidator Grant \textit{BlackHoleWeather} (101086804). FT thanks Elisabetta Liuzzo, Ranieri Baldi and Gabriele Bruni for useful discussions. The authors thank the referee for constructive comments. 

\section*{Data Availability}


The data underlying this article are available in the article.


\bibliographystyle{mnras}
\bibliography{example} 




\appendix

\section{Table A}

\begin{table*}
\caption{disc wind parameters of the RL sample from Tombesi et al.~(2014).\\ Notes: (1) Source name. (2) Observation number. (3) Logarithm of the ionization parameter. (4) Equivalent hydrogen column density. (5) Outflow velocity in units of the speed of light.}
\centering
\setlength{\tabcolsep}{4pt}
\begin{tabular}{lclll}
\hline\hline
(1)	& (2)		& (3)	& (4)		& (5)		\\
\noalign{\smallskip}
\hline
\noalign{\smallskip}
Source	&num		&$log\xi$	&$N_H$		&$v_{w}$		\\
\noalign{\smallskip}
\hline
\noalign{\smallskip}
		&		&erg cm/s	&$10^{22}$~cm$^{-2}$	&$c$		\\
\noalign{\smallskip}
\hline
\noalign{\smallskip}
4C+74.26	&1		&4.62$\pm0.25$ 		&$>$4				&0.045$\pm0.008$	\\
&2		&4.06$\pm0.45$		&$>$0.6				&0.185$\pm0.026$		\\
3C 120		&3		&3.80$\pm0.20$		&$1.1^{+0.5}_{-0.4}	$	&0.076	$\pm0.003$	\\
&4		&4.91$\pm1.03$		&$>$2				&0.161	$\pm0.006$	\\
PKS 1549-79	&5		&4.91$\pm0.49$		&$>$14				&0.276$\pm0.006$	\\
&6		&4.91$\pm0.49$		&$>$14				&0.427	$\pm0.005$	\\
3C 105		&7		&3.81$	\pm1.30$		&$>$2				&0.227$\pm0.033$	\\
3C 390.3	&8		&$5.60^{+0.20}_{-0.80}$	&$>$3				&0.146$\pm0.004$	\\
3C 111		&9		&5.00$\pm0.30$		&$>$20				&0.041	$\pm0.003$	\\
&10		&4.32$\pm0.12$		&7.7$\pm2.9$			&0.106$\pm0.006$	\\
3C 445		&11		&$1.42^{+0.13}_{-0.08}$	&$18.5^{+0.6}_{-0.7}$	&0.034$\pm0.001$	\\
\hline
\end{tabular}
\end{table*}

\begin{table*}
\caption{Main parameters of the sources and observations with disc winds for Tombesi et al.~(2014).\\ Notes: (1) Observation number. (2) Logarithm of the SMBH mass. (3) Logarithm of the X-ray luminosity in the energy range E$=$2--10 keV. (4) Logarithm of the Schwarschild radius. (5) Logarithm of the Eddington luminosity. (6) Logarithm of the Eddington mass accretion rate. (7) Logarithm of the bolometric luminosity. (8) Logarithm of the radiation momentum rate. (9) Logarithm of the mass accretion rate. (10) Logarithm of the Eddington ratio. }
\centering
\setlength{\tabcolsep}{4pt}
\begin{tabular}{lccccccccc}
\hline\hline
(1)	& (2)	& (3)	& (4) & (5)	& (6) & (7) & (8) & (9) & (10)\\
\noalign{\smallskip}
\hline
\noalign{\smallskip}
num	& $M_{BH}$ & $logL_X$ &log$r_{S}$		&log$L_{Edd}$		&log$\dot{M}_{Edd}$			&log$L_{bol}$			&log$\dot{p}_{rad}$		&log$\dot{M}_{acc}$ & $log\lambda$\\
\noalign{\smallskip}
\hline
\noalign{\smallskip}
	& $M_{\odot}$ & erg/s	&cm		&erg/s			&$M_{\odot}/yr$					&erg/s				& erg/cm 					&$M_{\odot}/yr$ & \\
\noalign{\smallskip}
\hline
\noalign{\smallskip}
1 & 9.6$\pm$0.5	&	44.60	&	15.07	&	47.71	&	1.96	&	46.30	&	35.82	&	0.55	&	-1.41\\
2 & 9.6$\pm$0.5	&	44.71	&	15.07	&	47.71	&	1.96	&	46.30	&	35.82 &	0.55	&	-1.41\\
3 & 7.7$^{+0.3}_{-0.2}$	&	43.79	&	13.17	&	45.81	&	0.06	&	45.33	&	34.85	&	-0.42	&	-0.48\\
4 & 7.7$^{+0.3}_{-0.2}$	&	43.88	&	13.17	&	45.81	&	0.06	&	45.33	&	34.85	&	-0.42	&	-0.48\\
5 & 8.1$\pm$0.3	&	44.46	&	13.57	&	46.21	&	0.46	&	45.78	&	35.30 &	0.02	&	-0.43\\
6 & 8.1$\pm$0.3	&	44.46	&	13.57	&	46.21	&	0.46	&	45.78	&	35.30	&	0.02	&	-0.43\\
7 & 7.8$\pm$0.5	&	43.90	&	13.27	&	45.91	&	0.16	&	45.17	&	34.69	&	-0.59	&	-0.75\\
8 & 8.8$^{+0.2}_{-0.6}$	&	44.18	&	14.27	&	46.91	&	1.16	&	45.31	&	34.83	&	-0.44	&	-1.60\\
9 & 8.1$\pm$0.5	&	43.87	&	13.57	&	46.21	&	0.46	&	44.68	&	34.20	&	-1.07	&	-1.53\\
10 & 8.1$\pm$0.5	&	44.32	&	13.57	&	46.21	&	0.46	&	44.68	&	34.20	&	-1.07	&	-1.53\\
11 & 8.3$\pm$0.3	&	43.85	&	13.77	&	46.41	&	0.66	&	45.10	&	34.63	&	-0.65	&	-1.31\\
\hline
\end{tabular}
\end{table*}

\begin{table*}
\caption{Estimated disc winds parameters for the sample of Tombesi et al.~(2014).\\Notes: (1) Observation number. (2) Logarithm of the ionizing luminosity in the 1-1000 Ryd band. (3)-(4) Logarithm of the minimum (maximum) distance of the wind from the source. (5)-(6) Logarithm of the minimum (maximum) mass outflow rate. (7)-(8) Logarithm of the minimum (maximum) wind kinetic power.}
\centering
\setlength{\tabcolsep}{4pt}
\begin{tabular}{lccccccc}
\hline\hline
(1)	& (2) & (3)	& (4) & (5)	& (6) & (7) & (8)\\
\noalign{\smallskip}
\hline
\noalign{\smallskip}
num		& $logL_{ion}$ 		& $logr_{min}$		& $logr_{max}$		& $log\dot{M}_{w}^{min}$		& $log\dot{M}_{w}^{max}$		& $logL_{w}^{min}$		& $logL_{w}^{max}$ \\
\noalign{\smallskip}
\hline
\noalign{\smallskip}
		& erg/s		& cm		&cm		&$M_{\odot}/yr$		&$M_{\odot}/yr$		& erg/s		& erg/s\\		
\noalign{\smallskip}
\hline
\noalign{\smallskip}
1 & 45.48	&	17.76	&	18.26	&	0.63	&	1.12	&	44.39	&	44.88\\
2 & 45.58	&	16.54	&	19.75	&	-0.81	&	2.40	&	44.18	&	47.39\\
3 & 44.66	&	15.41	&	18.82	&	-2.06	&	1.35	&	42.15	&	45.57\\
4 & 44.75	&	14.76	&	17.54	&	-2.13	&	0.66	&	42.74	&	45.52\\
5 & 45.34	&	14.69	&	17.28	&	-1.12	&	1.48	&	44.22	&	46.81\\
6 & 45.34	&	14.31	&	17.28	&	-1.31	&	1.67	&	44.41	&	47.38\\
7 & 44.78	&	14.56	&	18.67	&	-2.18	&	1.94	&	42.99	&	47.10\\
8 & 45.05	&	15.94	&	16.98	&	-0.81	&	0.23	&	43.97	&	45.01\\
9 & 44.75	&	16.34	&	16.45	&	-0.13	&	-0.03	&	43.55	&	43.65\\
10 & 45.20	&	15.52	&	17.99	&	-0.96	&	1.51	&	43.54	&	46.02\\
11 & 44.72	&	16.71	&	20.04	&	0.11	&	3.44	&	43.63	&	46.96\\
\hline
\end{tabular}
\end{table*}

\begin{table*}
\caption{Normalized disc winds parameters for the sample of Tombesi et al.~(2014). Notes: (1) Observation number. (2)-(3) Logarithm of the minimum (maximum) momentum rate of the wind. (4)-(5) Logarithm of the minimum (maximum) distance of the wind normalized to the Schwarschild radius. (6)-(7) Logarithm of the minimum (maximum) wind kinetic power normalized to the Eddington luminosity.}
\centering
\begin{tabular}{lcccccc}
\hline\hline
(1)	& (2) & (3)	& (4) & (5)	& (6) & (7)\\
\noalign{\smallskip}
\hline
\noalign{\smallskip}
num		& $log\dot{p}^{min}_{out}$		& $log\dot{p}^{max}_{out}$	& $log(\frac{r_{min}}{r_S})$	  & $log(\frac{r_{max}}{r_S})$	& $log(\frac{L^{min}_{out}}{L_{Edd}})$		& $log(\frac{L^{max}_{out}}{L_{Edd}})$\\
\noalign{\smallskip}
\hline
\noalign{\smallskip}
		& erg/cm	&erg/cm	&	&	&	&\\	
\noalign{\smallskip}
\hline
\noalign{\smallskip}
1 & 35.56	&	36.05	&	2.69	&	3.19	&	-3.33	&	-2.83\\
2 & 34.73	&	37.95	&	1.47	&	4.68	&	-3.54	&	-0.33\\
3 & 33.10	&	36.51	&	2.24	&	5.65	&	-3.66	&	-0.25\\
4 & 33.36	&	36.14	&	1.59	&	4.37	&	-3.07	&	-0.29\\
5 & 34.60	&	37.20	&	1.12	&	3.71	&	-2.00	&	0.60\\
6 & 34.60	&	37.58	&	0.74	&	3.71	&	-1.81	&	1.17\\
7 & 33.46	&	37.57	&	1.29	&	5.40	&	-2.93	&	1.19\\
8 & 34.63	&	35.67	&	1.67	&	2.71	&	-2.94	&	-1.91\\
9 & 34.76	&	34.86	&	2.77	&	2.88	&	-2.67	&	-2.57\\
10 & 34.34	&	36.82	&	1.95	&	4.42	&	-2.67	&	-0.20\\
11 & 34.92	&	38.25	&	2.94	&	6.27	&	-2.78	&	0.55\\
\hline
\end{tabular}
\end{table*}

\begin{table*}
\caption{Normalized disc wind mass and momentum rates for the sample of Tombesi et al.~(2014). Notes: (1) Observation number. (2)-(3) Logarithm of the ratio between the minimum (maximum) mass outflow rate and the accretion rate. (4)-(5) Logarithm of the ratio between the minimum (maximum) momentum rate of the outflow and the momentum rate of the radiation.}
\centering
\begin{tabular}{lcccc}
\hline\hline
(1)	& (2) & (3)	& (4) & (5)	\\
\noalign{\smallskip}
\hline
\noalign{\smallskip}
num		&$log(\frac{\dot{M}_{out}^{min}}{\dot{M}_{acc}})$	&$log(\frac{\dot{M}_{out}{max}}{\dot{M}_{acc}})$		& $log(\frac{\dot{p}_{out}{min}}{\dot{p}_{rad}})$	& $log(\frac{\dot{p}_{out}{max}}{\dot{p}_{rad}})$		\\
\noalign{\smallskip}
\hline
\noalign{\smallskip}
1 & 0.08	&	0.58	&	-0.27	&	0.23\\
2 & -1.36	&	1.86	&	-1.09	&	2.12\\
3 & -1.64	&	1.78	&	-1.76	&	1.66\\
4 & -1.70	&	1.08	&	-1.50	&	1.29\\
5 & -1.14	&	1.45	&	-0.70	&	1.89\\
6 & -1.33	&	1.64	&	-0.70	&	2.27\\
7 & -1.59	&	2.52	&	-1.23	&	2.88\\
8 & -0.36	&	0.67	&	-0.20	&	0.84\\
9 & 0.94	&	1.04	&	0.55	&	0.66\\
10 & 0.11	&	2.59	&	0.14	&	2.61\\
11 & 0.77	&	4.09	&	0.30	&	3.63\\

\hline
\end{tabular}
\end{table*}

\begin{table*}
\caption{\label{t7}disc wind parameters of the heterogeneous sample from Gofford et al.~(2015). Notes: (1)Source name. (2) Observation number. (3) Logarithm of the ionization parameter. (4) Equivalent hydrogen column density. (5) Outflow velocity in units of the speed of light.}
\centering
\setlength{\tabcolsep}{4pt}
\begin{tabular}{lclll}
\hline\hline
(1)	& (2)		& (3)	& (4)		& (5)		\\
\noalign{\smallskip}
\hline
\noalign{\smallskip}
Source	&num		&$log\xi$	&$N_H$		&$v_{w}$		\\
\noalign{\smallskip}
\hline
\noalign{\smallskip}
		&		&erg cm/s	&$10^{22}$~cm$^{-2}$	&$c$		\\
\noalign{\smallskip}
\hline
\noalign{\smallskip}
3C 111			&12		&$4.45^{+0.40}_{-0.46}$		& $10.0^{+9.0}_{-18.}$		&$0.072^{+0.041}_{-0.038}$	\\
3C 390.3		&13		&$>$5.46				&$>$50.12	&0.145$\pm0.007$			\\
4C +74.26		&14		&4.06$\pm0.45$			&$>$0.63	&0.185$\pm0.026$			\\
ESO 103-G035	&15		&4.36$\pm1.19$			&$>$0.79	&0.056$\pm0.025$			\\	
MR 2251-178		&16		&3.26$\pm0.12$			&0.32$\pm$0.15		&0.137$\pm0.008$			\\
Mrk 279		&17		&$4.42^{+0.15}_{-0.27}$		&25.12$\pm$11.57	&0.220$\pm0.006$			\\
Mrk 766		&18		&$3.86^{+0.37}_{-0.25}$		&5.01$\pm$2.31		&$0.039^{+0.030}_{-0.026}$	\\
NGC 4051		&19		&$3.50^{+0.53}_{-0.50}$		&6.31$\pm$1.45		&$0.018^{+0.004}_{-0.005}$	\\
NGC 4151		&20		&3.69$\pm0.64$			&$>$0.50	&0.055$\pm0.023$			\\
NGC 5506		&21		&$5.04^{+0.29}_{-0.17}$		&15.85$\pm$10.95		&0.246$\pm0.006$			\\
PDS 456		&22		&$4.06^{+0.28}_{-0.15}$		&10.00$\pm$2.30		&0.273$\pm0.028$			\\
SW J2127		&23		&$4.16^{+0.29}_{-0.13}$		&6.31$\pm$4.36		&0.231$\pm0.006$			\\
\hline
\end{tabular}
\end{table*}

\begin{table*}
\caption{Main parameters of the sources and observations with disc winds for Gofford et al.~(2015). Notes: (1) Observation number. (2) Logarithm of the SMBH mass. (3) Logarithm of the X-ray luminosity in the energy range E$=$2--10 keV. (4) Logarithm of the Schwarschild radius. (5) Logarithm of the Eddington luminosity. (6) Logarithm of the Eddington mass accretion rate. (7) Logarithm of the bolometric luminosity. (8) Logarithm of the radiation momentum rate. (9) Logarithm of the mass accretion rate. (10) Logarithm of the Eddington ratio.}
\centering
\setlength{\tabcolsep}{4pt}
\begin{tabular}{lccccccccc}
\hline\hline
(1)	& (2)	& (3)	& (4) & (5)	& (6) & (7) & (8) & (9) & (10)\\
\noalign{\smallskip}
\hline
\noalign{\smallskip}
num	& $M_{BH}$ & $logL_X$ &log$r_{S}$		&log$L_{Edd}$		&log$\dot{M}_{Edd}$			&log$L_{bol}$			&log$\dot{p}_{rad}$		&log$\dot{M}_{acc}$ & $log\lambda$\\
\noalign{\smallskip}
\hline
\noalign{\smallskip}
	& $M_{\odot}$ & erg/s	&cm		&erg/s			&$M_{\odot}/yr$					&erg/s				& erg/cm 					&$M_{\odot}/yr$ & \\
\noalign{\smallskip}
\hline
\noalign{\smallskip}
12 & 8.1$\pm$0.5	&	44.35	&	13.57	&	46.21	&	0.46	&	44.68	&	34.20	&	-1.07	&	-1.53\\
13 & 8.8$^{+0.2}_{-0.6}$	&	44.42	&	14.27	&	46.91	&	1.16	&	45.31	&	34.83	&	-0.44	&	-1.60\\
14 & 9.6	&	44.95	&	15.07	&	47.71	&	1.96	&	46.30	&	35.82	&	0.55	&	-1.41\\
15 & 7.4$\pm$0.1	&	43.47	&	12.87	&	45.51	&	-0.24	&	44.70	&	34.22	&	-1.06	&	-0.82\\
16 & 8.7$\pm$0.1&	44.61	&	14.17	&	46.81	&	1.06	&	45.78	&	35.30	&	0.03	&	-1.03\\
17 & 7.5$\pm$0.1	&	42.78	&	12.97	&	45.61	&	-0.14	&	44.87	&	34.39	&	-0.88	&	-0.74\\
18 & 6.2$^{+0.3}_{-0.6}$	&	42.73	&	11.67	&	44.31	&	-1.44	&	43.93	&	33.46	&	-1.82	&	-0.38\\
19 & 6.3$\pm$0.2	&	41.65	&	11.77	&	44.41	&	-1.34	&	43.02	&	32.54	&	-2.73	&	-1.39\\
20 & 7.1$\pm$0.1	&	42.34	&	12.57	&	45.21	&	-0.54	&	43.83	&	33.35	&	-1.92	&	-1.38\\
21 & 7.3$\pm$0.7	&	43.19	&	12.77	&	45.41	&	-0.34	&	44.10	&	33.62	&	-1.65	&	-1.31\\
22 & 9.4$\pm$0.3	&	44.62	&	14.87	&	47.51	&	1.76	&	46.30	&	35.82	&	0.55	&	-1.21\\
23 & 7.2	&	43.15	&	12.67	&	45.31	&	-0.44	&	44.54	&	34.06	&	-1.21	&	-0.77\\
\hline
\end{tabular}
\end{table*}

\begin{table*}
\caption{Estimated disc winds parameters for the sample of Gofford et al.~(2015). Notes: (1) Observation number. (2) Logarithm of the ionizing luminosity in the 1-1000 Ryd band. (3)-(4) Logarithm of the minimum (maximum) distance of the wind from the source. (5)-(6) Logarithm of the minimum (maximum) mass outflow rate. (7)-(8) Logarithm of the minimum (maximum) wind kinetic power.}
\centering
\setlength{\tabcolsep}{4pt}
\begin{tabular}{lccccccc}
\hline\hline
(1)	& (2) & (3)	& (4) & (5)	& (6) & (7) & (8)\\
\noalign{\smallskip}
\hline
\noalign{\smallskip}
num		& $logL_{ion}$ 		& $logr_{min}$		& $logr_{max}$		& $log\dot{M}_{w}^{min}$		& $log\dot{M}_{w}^{max}$		& $log(L_{w}^{min})$		& $log(L_{w}^{max})$ \\
\noalign{\smallskip}
\hline
\noalign{\smallskip}
		& erg/s		& cm		&cm		&$M_{\odot}/yr$		&$M_{\odot}/yr$		& erg/s		& erg/s\\		
\noalign{\smallskip}
\hline
\noalign{\smallskip}
12 & 44.70	&	15.86	&	17.25	&	-0.68	&	0.72	&	43.49	&	44.88\\
13 & 44.70	&	15.95	&	15.54	&	0.42	&	0.01	&	45.19	&	44.79\\
14 & 47.00	&	16.54	&	21.14	&	-0.79	&	3.82	&	44.20	&	48.80\\
15 & 43.90	&	15.37	&	17.64	&	-2.37	&	-0.10	&	41.58	&	43.85\\
16 & 45.20	&	15.90	&	20.44	&	-1.86	&	2.69	&	42.87	&	47.41\\
17 & 43.40	&	14.28	&	15.58	&	-1.36	&	-0.07	&	43.77	&	45.07\\
18 & 43.20	&	14.49	&	16.64	&	-2.61	&	-0.46	&	41.02	&	43.18\\
19 & 42.50	&	15.26	&	16.20	&	-2.08	&	-1.14	&	40.89	&	41.83\\
20 & 42.90	&	15.09	&	17.51	&	-2.86	&	-0.44	&	41.07	&	43.49\\
21 & 43.70	&	13.99	&	15.46	&	-1.81	&	-0.34	&	43.42	&	44.89\\
22 & 45.30	&	16.00	&	18.24	&	0.04	&	2.29	&	45.37	&	47.61\\
23 & 43.70	&	13.94	&	16.74	&	-2.28	&	0.51	&	42.90	&	45.69\\

\hline
\end{tabular}
\end{table*}

\begin{table*}
\caption{Normalized disc winds parameters for the sample of Gofford et al.~(2015). Notes: (1) Observation number. (2)-(3) Logarithm of the minimum (maximum) momentum rate of the wind. (4)-(5) Logarithm of the minimum (maximum) distance of the wind normalized to the Schwarschild radius. (6)-(7) Logarithm of the minimum (maximum) wind kinetic power normalized to the Eddington luminosity.}
\centering
\begin{tabular}{lcccccc}
\hline\hline
(1)	& (2) & (3)	& (4) & (5)	& (6) & (7) \\
\noalign{\smallskip}
\hline
\noalign{\smallskip}
num		& $log\dot{p}^{min}_{out}$		& $log\dot{p}^{max}_{out}$	& $log(\frac{r_{min}}{r_S})$	  & $log(\frac{r_{max}}{r_S})$	& $log(\frac{L^{min}_{out}}{L_{Edd}})$		& $log(\frac{L^{max}_{out}}{L_{Edd}})$\\
\noalign{\smallskip}
\hline
\noalign{\smallskip}
		& erg/cm	&erg/cm	&	& 	&	&\\	
\noalign{\smallskip}
\hline
\noalign{\smallskip}
12 & 34.46	&	35.85	&	2.29	&	3.68	&	-2.72	&	-1.33\\
13 & 35.86	&	35.45	&	1.68	&	1.27	&	-1.72	&	-2.13\\
14 & 34.76	&	39.36	&	1.47	&	6.07	&	-3.51	&	1.09\\
15 & 32.66	&	34.92	&	2.50	&	4.77	&	-3.93	&	-1.67\\
16 & 33.56	&	38.10	&	1.73	&	6.27	&	-3.95	&	0.60\\
17 & 34.26	&	35.55	&	1.32	&	2.61	&	-1.84	&	-0.54\\
18 & 32.26	&	34.41	&	2.82	&	4.97	&	-3.29	&	-1.14\\
19 & 32.46	&	33.40	&	3.49	&	4.43	&	-3.53	&	-2.59\\
20 & 32.16	&	34.58	&	2.52	&	4.94	&	-4.14	&	-1.72\\
21 & 33.86	&	35.33	&	1.22	&	2.69	&	-1.99	&	-0.52\\
22 & 35.76	&	38.00	&	1.13	&	3.37	&	-2.15	&	0.10\\
23 & 33.36	&	36.15	&	1.27	&	4.07	&	-2.42	&	0.38\\
\hline
\end{tabular}
\end{table*}

\begin{table*}
\caption{Normalized disc wind mass and momentum rates for the sample of Gofford et al.~(2015). Notes: (1) Observation number. (2)-(3) Logarithm of the ratio between the minimum (maximum) mass outflow rate and the accretion rate. (4)-(5) Logarithm of the ratio between the minimum (maximum) momentum rate of the outflow and the momentum rate of the radiation.}
\centering
\begin{tabular}{lcccc}
\hline\hline
(1)	& (2) & (3)	& (4) & (5)	\\
\noalign{\smallskip}
\hline
\noalign{\smallskip}
num		&$log(\frac{\dot{M}_{out}^{min}}{\dot{M}_{acc}})$	&$log(\frac{\dot{M}_{out}{max}}{\dot{M}_{acc}})$		& $log(\frac{\dot{p}_{out}{min}}{\dot{p}_{rad}})$	& $log(\frac{\dot{p}_{out}{max}}{\dot{p}_{rad}})$		\\
\noalign{\smallskip}
\hline
\noalign{\smallskip}
12 & 0.40	&	1.79	&	0.25	&	1.65\\
13 & 0.86	&	0.45	&	1.02	&	0.62\\
14 & -1.33	&	3.27	&	-1.07	&	3.54\\
15 & -1.31	&	0.95	&	-1.57	&	0.70\\
16 & -1.88	&	2.66	&	-1.75	&	2.80\\
17 & -0.48	&	0.82	&	-0.14	&	1.16\\
18 & -0.79	&	1.36	&	-1.20	&	0.95\\
19 & 0.66	&	1.60	&	-0.09	&	0.85\\
20 & -0.94	&	1.48	&	-1.20	&	1.22\\
21 & -0.16	&	1.31	&	0.23	&	1.70\\
22 & -0.50	&	1.74	&	-0.07	&	2.18\\
23 & -1.07	&	1.73	&	-0.71	&	2.09\\
\hline
\end{tabular}
\end{table*}

\begin{table*}
\caption{\label{t8}disc wind parameters of the RQ sample from Tombesi et al.~(2012, 2013). Notes: (1)Source name. (2) Observation number. (3) Logarithm of the ionization parameter. (4) Equivalent hydrogen column density. (5) Outflow velocity in units of the speed of light.}
\centering
\setlength{\tabcolsep}{4pt}
\begin{tabular}{lclll}
\hline\hline
(1)	& (2)		& (3)	& (4)		& (5)		\\
\noalign{\smallskip}
\hline
\noalign{\smallskip}
Source	&num		&$log\xi$	&$N_H$		&$v_{w}$		\\
\noalign{\smallskip}
\hline
\noalign{\smallskip}
		&		&erg cm/s	&$10^{22}$~cm$^{-2}$	&$c$		\\
\noalign{\smallskip}
\hline
\noalign{\smallskip}
NGC 4151		&24		&$4.41^{+0.92}_{-0.08}$		&$>$2		&0.106$\pm0.007$		\\
IC 4329A		&25		&5.34$\pm0.94$		&$>$2		&0.098$\pm0.004$		\\
Ark 120		&26		&4.55$\pm1.29$		&$>$0.7	&0.287$\pm0.022$		\\
Mrk 79			&27		&4.19$\pm0.23$		&19.4$\pm12.0$	&0.092$\pm0.004$		\\
NGC 4051		&28		&4.37$\pm1.54$		&$>$0.80		&0.037$\pm0.025$		\\
			&29 	&$3.06^{+0.17}_{-0.19}$		&$5.2^{+1.1}_{-1.3}$	&0.202$\pm0.006$		\\
Mrk 766		&30		&$3.46^{+0.71}_{-0.32}$		&$2.1^{+0.4}_{-0.3}$	&0.082$\pm0.006$	\\
			&31		&$4.28^{+0.08}_{-0.11}$		&$10.0^{+6.4}_{-3.5}$	&0.088$\pm0.002$		\\
Mrk 841		&32		&3.91$\pm1.24$		&$>$1		&0.055$\pm0.025$		\\
1H0419-577		&33		&3.69$\pm0.53$		&17.3$\pm14.3$	&0.079$\pm0.007$		\\
Mrk 290		&34		&3.91$\pm1.17$		&25.9$\pm24.0$	&0.163$\pm0.024$		\\
Mrk 205		&35		&4.92$\pm0.56$		&$>$14.5		&0.100$\pm0.004$		\\
PG 1211+143		&36		&$2.87^{+0.12}_{-0.10}$		&$8.0^{+2.2}_{-1.1}$		&0.151$\pm0.003$		\\
MCG-5-23-16		&37		&4.33$\pm0.08$	&4.0$\pm1.2$		&0.116$\pm0.004$		\\
NGC 4507		&38		&4.53$\pm1.15$		&$>$0.9		&0.199$\pm0.024$		\\
NGC 7582		&39		&$3.39^{+0.09}_{-0.15}$		&$23.4^{+21.1}_{-9.8}$	&0.285$\pm0.003$		\\
\hline
\end{tabular}
\end{table*}

\begin{table*}
\caption{Main parameters of the sources and observations with disc winds for Tombesi et al.~(2012, 2013). Notes: (1) Observation number. (2) Logarithm of the SMBH mass. (3) Logarithm of the X-ray luminosity in the energy range E$=$4--10 keV. (4) Logarithm of the Schwarschild radius. (5) Logarithm of the Eddington luminosity. (6) Logarithm of the Eddington mass accretion rate. (7) Logarithm of the bolometric luminosity. (8) Logarithm of the radiation momentum rate. (9) Logarithm of the mass accretion rate. (10) Logarithm of the Eddington ratio.}
\centering
\setlength{\tabcolsep}{4pt}
\begin{tabular}{lccccccccc}
\hline\hline
(1)	& (2)	& (3)	& (4) & (5)	& (6) & (7) & (8) & (9) & (10)\\
\noalign{\smallskip}
\hline
\noalign{\smallskip}
num	& $M_{BH}$ & $logL_X$ &log$r_{S}$		&log$L_{Edd}$		&log$\dot{M}_{Edd}$			&log$L_{bol}$			&log$\dot{p}_{rad}$		&log$\dot{M}_{acc}$ & $log\lambda$\\
\noalign{\smallskip}
\hline
\noalign{\smallskip}
	& $M_{\odot}$ & erg/s	&cm		&erg/s			&$M_{\odot}/yr$					&erg/s				& erg/cm 					&$M_{\odot}/yr$ & \\
\noalign{\smallskip}
\hline
\noalign{\smallskip}
24 & 7.1$\pm$0.2	&	42.50	&	12.57	&	45.21	&	-0.54	&	43.83	&	33.35	&	-1.92	&	-1.38\\
25 & 8.1$\pm$0.2	&	43.70	&	13.57	&	46.21	&	0.46	&	45.00	&	34.52	&	-0.75	&	-1.21\\
26 & 8.2$\pm$0.1	&	44.00	&	13.67	&	46.31	&	0.56	&	45.38	&	34.90	&	-0.37	&	-0.93\\
27 & 7.7$\pm$0.1	&	43.40	&	13.17	&	45.81	&	0.06	&	44.57	&	34.09	&	-1.18	&	-1.24\\
28 & 6.3$\pm$0.4	&	41.50	&	11.77	&	44.41	&	-1.34	&	43.02	&	32.54	&	-2.73	&	-1.39\\
29 & 6.3$\pm$0.4	&	41.00	&	11.77	&	44.41	&	-1.34	&	43.02	&	32.54	&	-2.73	&	-1.39\\
30 & 6.1$\pm$0.4	&	42.60	&	11.57	&	44.21	&	-1.54	&	44.30	&	33.82	&	-1.45	&	0.09\\
31 & 6.1$\pm$0.4	&	42.80	&	11.57	&	44.21	&	-1.54	&	44.30	&	33.82	&	-1.45	&	0.09\\
32 & 7.8$\pm$0.5	&	43.50	&	13.27	&	45.91	&	0.16	&	45.84	&	35.36	&	0.09	&	-0.07\\
33 & 8.6$\pm$0.5	&	44.30	&	14.07	&	46.71	&	0.96	&	45.95	&	35.47	&	0.20	&	-0.76\\
34 & 7.7$\pm$0.5	&	43.20	&	13.17	&	45.81	&	0.06	&	44.57	&	34.09	&	-1.18	&	-1.24\\
35 & 8.6$\pm$1.0	&	43.80	&	14.07	&	46.71	&	0.96	&	45.05	&	34.58	&	-0.70	&	-1.66\\
36 & 8.2$\pm$0.2	&	43.70	&	13.67	&	46.31	&	0.56	&	45.81	&	35.33	&	0.06	&	-0.50\\
37 & 7.6$\pm$1.0	&	43.10	&	13.07	&	45.71	&	-0.04	&	44.31	&	33.83	&	-1.44	&	-1.40\\
38 & 6.4$\pm$0.5	&	43.10	&	11.87	&	44.51	&	-1.24	&	44.40	&	33.92	&	-1.35	&	-0.11\\
39 & 7.1$\pm$1.0	&	41.60	&	12.57	&	45.21	&	-0.54	&	43.30	&	32.82	&	-2.45	&	-1.91\\
\hline
\end{tabular}
\end{table*}

\begin{table*}
\caption{Estimated disc winds parameters for the sample of Tombesi et al.~(2012, 2013). Notes: (1) Observation number. (2) Logarithm of the ionizing luminosity in the 1-1000 Ryd band. (3)-(4) Logarithm of the minimum (maximum) distance of the wind from the source. (5)-(6) Logarithm of the minimum (maximum) mass outflow rate. (7)-(8) Logarithm of the minimum (maximum) wind kinetic power.}
\centering
\setlength{\tabcolsep}{4pt}
\begin{tabular}{lccccccc}
\hline\hline
(1)	& (2) & (3)	& (4) & (5)	& (6) & (7) & (8)\\
\noalign{\smallskip}
\hline
\noalign{\smallskip}
num		& $logL_{ion}$ 		& $logr_{min}$		& $logr_{max}$		& $log\dot{M}_{w}^{min}$		& $log\dot{M}_{w}^{max}$		& $log(L_{w}^{min})$		& $log(L_{w}^{max})$ \\
\noalign{\smallskip}
\hline
\noalign{\smallskip}
		& erg/s		& cm		&cm		&$M_{\odot}/yr$		&$M_{\odot}/yr$		& erg/s		& erg/s\\		
\noalign{\smallskip}
\hline
\noalign{\smallskip}
24 & 42.90	&	14.52	&	16.19	&	-2.55	&	-0.88	&	41.96	&	43.63\\
25 & 44.10	&	15.59	&	16.46	&	-1.51	&	-0.64	&	42.92	&	43.80\\
26 & 44.50	&	14.75	&	18.10	&	-2.33	&	1.02	&	43.03	&	46.39\\
27 & 43.90	&	15.24	&	16.42	&	-0.90	&	0.28	&	43.48	&	44.66\\
28 & 42.30	&	14.63	&	16.03	&	-3.29	&	-1.89	&	40.30	&	41.70\\
29 & 42.00	&	13.16	&	16.22	&	-3.21	&	-0.15	&	41.85	&	44.92\\
30 & 43.20	&	13.74	&	17.42	&	-3.41	&	0.26	&	40.87	&	44.54\\
31 & 43.40	&	13.68	&	16.12	&	-2.77	&	-0.33	&	41.58	&	44.02\\
32 & 43.90	&	15.79	&	17.99	&	-1.86	&	0.34	&	42.07	&	44.27\\
33 & 44.60	&	16.27	&	17.67	&	0.02	&	1.42	&	44.27	&	45.66\\
34 & 43.60	&	14.75	&	16.28	&	-1.02	&	0.51	&	43.86	&	45.39\\
35 & 44.20	&	16.07	&	16.12	&	-0.16	&	-0.11	&	44.29	&	44.34\\
36 & 44.30	&	15.31	&	18.53	&	-1.00	&	2.22	&	43.81	&	47.03\\
37 & 43.50	&	14.94	&	16.57	&	-1.78	&	-0.16	&	42.80	&	44.43\\
38 & 43.50	&	13.27	&	17.02	&	-3.87	&	-0.12	&	41.18	&	44.93\\
39 & 43.40	&	13.66	&	16.64	&	-1.91	&	1.07	&	43.46	&	46.44\\
\hline
\end{tabular}
\end{table*}

\begin{table*}
\caption{Normalized disc winds parameters for the sample of Tombesi et al.~(2012, 2013). Notes: (1) Observation number. (2)-(3) Logarithm of the minimum (maximum) momentum rate of the wind. (4)-(5) Logarithm of the minimum (maximum) distance of the wind normalized to the Schwarschild radius. (6)-(7) Logarithm of the minimum (maximum) wind kinetic power normalized to the Eddington luminosity.}
\centering
\begin{tabular}{lcccccc}
\hline\hline
(1)	& (2) & (3)	& (4) & (5)	& (6) & (7) \\
\noalign{\smallskip}
\hline
\noalign{\smallskip}
num		& $log\dot{p}^{min}_{out}$		& $log\dot{p}^{max}_{out}$	& $log(\frac{r_{min}}{r_S})$	  & $log(\frac{r_{max}}{r_S})$	& $log(\frac{L^{min}_{out}}{L_{Edd}})$		& $log(\frac{L^{max}_{out}}{L_{Edd}})$\\
\noalign{\smallskip}
\hline
\noalign{\smallskip}
		& erg/cm	&erg/cm	&	&	&	&\\	
\noalign{\smallskip}
\hline
\noalign{\smallskip}
24 & 32.76	&	34.43	&	1.95	&	3.62	&	-3.26	&	-1.59\\
25 & 33.76	&	34.63	&	2.02	&	2.89	&	-3.29	&	-2.42\\
26 & 33.40	&	36.75	&	1.08	&	4.44	&	-3.28	&	0.07\\
27 & 34.34	&	35.52	&	2.07	&	3.25	&	-2.33	&	-1.15\\
28 & 31.56	&	32.95	&	2.86	&	4.26	&	-4.11	&	-2.72\\
29 & 32.37	&	35.44	&	1.39	&	4.45	&	-2.56	&	0.50\\
30 & 31.78	&	35.45	&	2.17	&	5.85	&	-3.35	&	0.33\\
31 & 32.46	&	34.89	&	2.11	&	4.55	&	-2.64	&	-0.20\\
32 & 33.16	&	35.36	&	2.52	&	4.72	&	-3.84	&	-1.64\\
33 & 35.19	&	36.59	&	2.20	&	3.60	&	-2.45	&	-1.05\\
34 & 34.47	&	36.00	&	1.58	&	3.11	&	-1.96	&	-0.43\\
35 & 35.12	&	35.17	&	2.00	&	2.05	&	-2.42	&	-2.37\\
36 & 34.46	&	37.67	&	1.64	&	4.86	&	-2.50	&	0.72\\
37 & 33.56	&	35.18	&	1.87	&	3.50	&	-2.92	&	-1.29\\
38 & 31.71	&	35.45	&	1.40	&	5.15	&	-3.33	&	0.41\\
39 & 33.82	&	36.81	&	1.09	&	4.07	&	-1.76	&	1.22\\
\hline
\end{tabular}
\end{table*}

\begin{table*}
\caption{Normalized disc wind mass and momentum rates for the sample of Tombesi et al.~(2012, 2013). Notes: (1) Observation number. (2)-(3) Logarithm of the ratio between the minimum (maximum) mass outflow rate and the accretion rate. (4)-(5) Logarithm of the ratio between the minimum (maximum) momentum rate of the outflow and the momentum rate of the radiation.}
\centering
\begin{tabular}{lcccc}
\hline\hline
(1)	& (2) & (3)	& (4) & (5)	\\
\noalign{\smallskip}
\hline
\noalign{\smallskip}
num		&$log(\frac{\dot{M}_{out}^{min}}{\dot{M}_{acc}})$	&$log(\frac{\dot{M}_{out}{max}}{\dot{M}_{acc}})$		& $log(\frac{\dot{p}_{out}{min}}{\dot{p}_{rad}})$	& $log(\frac{\dot{p}_{out}{max}}{\dot{p}_{rad}})$		\\
\noalign{\smallskip}
\hline
\noalign{\smallskip}
24 & -0.62	&	1.05	&	-0.60	&	1.07\\
25 & -0.76	&	0.11	&	-0.77	&	0.11\\
26 & -1.96	&	1.39	&	-1.50	&	1.85\\
27 & 0.29	&	1.47	&	0.25	&	1.43\\
28 & -0.55	&	0.84	&	-0.98	&	0.41\\
29 & -0.48	&	2.59	&	-0.17	&	2.89\\
30 & -1.96	&	1.72	&	-2.04	&	1.63\\
31 & -1.31	&	1.13	&	-1.37	&	1.07\\
32 & -1.95	&	0.25	&	-2.21	&	-0.01\\
33 & -0.18	&	1.22	&	-0.28	&	1.12\\
34 & 0.16	&	1.70	&	0.38	&	1.91\\
35 & 0.54	&	0.59	&	0.54	&	0.59\\
36 & -1.05	&	2.16	&	-0.87	&	2.34\\
37 & -0.34	&	1.29	&	-0.28	&	1.35\\
38 & -2.51	&	1.23	&	-2.21	&	1.53\\
39 & 0.55	&	3.53	&	1.00	&	3.98\\
\hline
\end{tabular}
\end{table*}

\section{Tables B}

\begin{table*}
\caption{Mean disc wind parameters for our RQ and RL sample. Notes: (1)Source name. (2) Observation number. (3) AGN' class. (4) Logarithm of the ionization parameter. (5) Equivalent hydrogen column density. (6) Outflow velocity in units of the speed of light. (7) Logarithm of the SMBH mass. (8) Logarithm of the X-ray luminosity in the energy range E$=$2--10 keV.}
\centering
\setlength{\tabcolsep}{4pt}
\begin{tabular}{lcclllll}
\hline\hline
(1)	& (2)		& (3)	& (4)		& (5) &(6) &(7) &(8)		\\
\noalign{\smallskip}
\hline
\noalign{\smallskip}
Source	&num & type		&$log\xi$	&$N_H$		&$v_{w}$ & $M_{BH}$ & $logL_X$		\\
\noalign{\smallskip}
\hline
\noalign{\smallskip}
		& &		&erg cm/s	&$10^{22}$~cm$^{-2}$	&$c$	& M$_{\odot}$ & erg/s	\\
\noalign{\smallskip}
\hline
\noalign{\smallskip}
4C+74.26	&	1	&	RL & 4.20$\pm$0.50	&	$>$1.47	&	0.150$\pm$0.004	&	9.6$\pm$0.5	&	44.78\\
3C 120	&	2	&	RL & 4.36$\pm$1.05	&	$>$1.55	&	0.120$\pm$0.004	&	7.7$^{+0.3}_{-0.2}$	&	43.79\\
PKS 1549-79	&	3	&	RL & 4.91$\pm$0.49	&	>14	&	0.351$\pm$0.006	&	8.1$\pm$0.3	&	44.46\\
3C 105	&	4	&	RL & 3.81$\pm$0.20	&	$>$2	&	0.230$\pm$0.003	&	7.8$\pm$0.5	&	43.90\\
3C 390.3	&	5	& RL &	$>$5.53	&	$>$26.56	&	0.146$\pm$0.008	&	8.8$^{+0.2}_{-0.6}$	&	44.30\\
3C 111	&	6	&	RL & 4.56$\pm$0.32	&	$>$11.93	&	0.070$\pm$0.008	&	8.1$\pm$0.5	&	44.11\\
3C 445	&	7	&	RL & 1.42$^{+0.13}_{-0.08}$	&	18.50$\pm$3.0	&	0.034$\pm$0.001	&	8.3$\pm$0.3	&	43.85\\
ESO 103-G035	&	8	&	RQ & 4.36$\pm$1.19	&	$>$0.79	&	0.056$\pm$0.025	&	7.4$\pm$0.1	&	43.47\\
MR 2251-178	&	9	&	RQ & 3.26$\pm$0.12	&	0.32$\pm$0.15	&	0.137$\pm$0.008	&	8.7$\pm$0.1	&	44.61\\
Mrk 279	&	10	&	RQ & 4.42$^{+0.15}_{-0.27}$	&	25.12$\pm$11.57	&	0.220$\pm$0.006	&	7.5$\pm$0.1	&	42.78\\
Mrk 766	&	11	&	RQ & 3.87$^{+0.71}_{-0.34}$	&	5.5$^{+6.4}_{-3.6}$	&	0.085$\pm$0.004	&	6.1$\pm$0.4	&	42.67\\
NGC 4051	&	12	& RQ &	3.61$\pm$1.55	&	4.65$\pm$1.51	&	0.085$\pm$0.025	&	6.3$\pm$0.4	&	41.58\\
NGC 4151	&	13	& RQ &	4.05$^{+0.94}_{-0.64}$	&	$>$1.25	&	0.081$\pm$0.023	&	7.1$\pm$0.2	&	42.42\\
NGC 5506	&	14	& RQ &	5.04$^{+0.29}_{-0.17}$	&	15.85$\pm$10.95	&	0.246$\pm$0.006	&	7.3$\pm$0.7	&	43.19\\
PDS 456	&	15	&	RQ & 4.06$^{+0.28}_{-0.15}$	&	10.00$\pm$2.30	&	0.273$\pm$0.028	&	9.4$\pm$0.3	&	44.62\\
SW J2127	&	16	&	RQ & 4.16$^{+0.29}_{-0.13}$	&	6.31$\pm$4.36	&	0.231$\pm$0.006	&	7.2$\pm$0.3	&	43.15\\
IC 4329A	&	17	&	RQ & 5.34$\pm$0.94	&	>2	&	0.106$\pm$0.007	&	8.1$\pm$0.2	&	43.70\\
Mrk 205	&	18	&	RQ & 4.92$\pm$0.56	&	$>$14.5	&	0.100$\pm$0.004	&	8.6$\pm$0.1	&	43.80\\
Ark 120	&	19	&	RQ & 4.55$\pm$1.29	&	$>$0.7	&	0.287$\pm$0.022	&	8.2$\pm$0.1	&	44.00\\
Mrk 79	&	20	&	RQ & 4.19$\pm$0.23	&	19.4$\pm$12.0	&	0.092$\pm$0.004	&	7.7$\pm$0.1	&	43.40\\
Mrk 841	&	21	&	RQ & 3.91$\pm$1.24	&	$>$1	&	0.055$\pm$0.025	&	7.8$\pm$0.5	&	43.50\\
1H0419-577	&	22	& RQ &	3.69$\pm$0.53	&	17.3$\pm$14.3	&	0.079$\pm$0.007	&	8.6$\pm$0.5	&	44.30\\
Mrk 290	&	23	&	RQ & 3.91$\pm$1.17	&	25.9$\pm$24.0	&	0.163$\pm$0.024	&	7.7$\pm$0.5	&	43.20\\
PG 1211+143	&	24	& RQ &	2.87$^{+0.12}_{-0.10}$	&	8.0$^{+2.2}_{-1.1}$ &	0.151$\pm$0.003	&	8.20$\pm$0.2   &	43.70\\
MCG-5-23-16	&	25	& RQ &	4.33$\pm$0.08	&	4.0$\pm$1.2	&	0.116$\pm$0.004	&	7.6$\pm$1.0	&	43.10\\
NGC 4507	&	26	& RQ &	4.53$\pm$1.15	&	$>$0.9	&	0.199$\pm$0.024	&	6.4$\pm$0.5	&	43.10\\
NGC 7582	&	27	& RQ & 	3.39$^{+0.09}_{-0.15}$	&	23.4$^{+21.1}_{-9.8}$	&	0.285$\pm$0.003	&	7.1$\pm$0.1	&	41.60\\
\hline
\end{tabular}
\end{table*}

\begin{table*}
\caption{Main parameters of the sources and observations with disc winds for our sample. Notes: (1) Observation number. (2) Logarithm of the Schwarschild radius. (3) Logarithm of the Eddington luminosity. (4) Logarithm of the Eddington mass accretion rate. (5) Logarithm of the bolometric luminosity (letters from $a$ to $f$ refer to the corresponding references: (a) \citet{Woo2002}, (b) \citet{Thorne2021}, (c) \citet{Gesù2013}, (d) \citet{Vasudevan2007}, (e) \citet{Reeves2021} and (f) \citet{Piotrovich2022}). (6) Logarithm of the radiation momentum rate. (7) Logarithm of the mass accretion rate. (8) Logarithm of the Eddington ratio. (9) Logarithm of the Radio Loudness.}
\centering
\setlength{\tabcolsep}{4pt}
\begin{tabular}{lcccccccc}
\hline\hline
(1)	& (2)		& (3)	& (4)		& (5) &(6) &(7)		&(8)&(9)\\
\noalign{\smallskip}
\hline
\noalign{\smallskip}
num	&log$r_{S}$		&log$L_{Edd}$		&log$\dot{M}_{Edd}$			&log$L_{bol}$			&log$\dot{p}_{rad}$		&log$\dot{M}_{acc}$ &  $log\lambda_{Edd}$ & $log$R$_x$\\
\noalign{\smallskip}
\hline
\noalign{\smallskip}
		&cm		&erg/s			&$M_{\odot}/yr$					&erg/s				& erg/cm 					&$M_{\odot}/yr$ & &\\
\noalign{\smallskip}
\hline
\noalign{\smallskip}
1 & 15.07	&	47.71	&	1.96	&	46.30$^a$	&	35.82	&	0.55	&	-1.41 & -2.99\\
2 & 13.17	&	45.81	&	0.06	&	45.33$^b$    &	34.85	&	-0.42	&	-0.48 & -3.08\\
3 & 13.57	&	46.21	&	0.46	&	45.78$^a$	&	35.30	&	0.03	&	-0.43 & --\\
4 & 13.27	&	45.91	&	0.16	&	45.17	&	34.69	&	-0.59	&	-0.75 & --\\
5 & 14.27	&	46.91	&	1.16	&	45.31$^b$	&	34.83	&	-0.44	&	-1.60 & -1.49\\
6 & 13.57	&	46.21	&	0.46	&	44.68	&	34.20	&	-1.07	&	-1.53 & -4.10\\
7 & 13.77	&	46.41	&	0.66	&	45.10	&	34.63	&	-0.65	&	-1.31 & -3.14\\
8 & 12.87	&	45.51	&	-0.24	&	44.70	&	34.22	&	-1.06	&	-0.82 & -4.37\\
9 & 14.17	&	46.81	&	1.06	&	45.78	&	35.30	&	0.03	&	-1.03 & -4.29\\
10 & 12.97	&	45.61	&	-0.14	&	44.87$^b$	&	34.39	&	-0.88	&	-0.74 & --\\
11 & 11.62	&	44.26	&	-1.49	&	44.12	&	33.64	&	-1.64	&	-0.15 & -3.40\\
12 & 11.77	&	44.41	&	-1.34	&	43.02$^b$	&	32.54	&	-2.73	&	-1.39 & -4.26\\
13 & 12.57	&	45.21	&	-0.54	&	43.83$^b$	&	33.35	&	-1.92	&	-1.38 & -3.38\\
14 & 12.77	&	45.41	&	-0.34	&	44.10	&	33.62	&	-1.65	&	-1.31 & -3.67\\
15 & 14.87	&	47.51	&	1.76	&	46.30$^e$	&	35.82	&	0.55	&	-1.21 & --\\
16 & 12.67	&	45.31	&	-0.44	&	44.54$^f$	&	34.06	&	-1.21	&	-0.77 & -4.56\\
17 & 13.57	&	46.21	&	0.46	&	45.00$^d$	&	34.52	&	-0.75	&	-1.21 & -4.02\\
18 & 14.07	&	46.71	&	0.96	&	45.05	&	34.58	&	-0.70	&	-1.66 & --\\
19 & 13.67	&	46.31	&	0.56	&	45.38$^b$	&	34.90	&	-0.37	&	-0.93 & --\\
20 & 13.17	&	45.81	&	0.06	&	44.57$^a$	&	34.09	&	-1.18	&	-1.24 & -4.10\\
21 & 13.27	&	45.91	&	0.16	&	45.84$^a$	&	35.36	&	0.09	&	-0.07 & --\\
22 & 14.07	&	46.71	&	0.96	&	45.95$^c$	&	35.47	&	0.20	&	-0.76 & --\\
23 & 13.17	&	45.81	&	0.06	&	44.57$^b$	&	34.09	&	-1.18	&	-1.24 & -4.07\\
24 & 13.67	&	46.31	&	0.56	&	45.77$^b$	&	35.33	&	0.06	&	-0.50 & --\\
25 & 13.07	&	45.71	&	-0.04	&	44.31	&	33.84	&	-1.44	&	-1.40 & -4.64\\
26 & 11.87	&	44.51	&	-1.24	&	44.40$^f$	&	33.92	&	-1.35	&	-0.11 & -3.69\\
27 & 12.57	&	45.21	&	-0.54	&	43.30$^d$	&	32.82	&	-2.45	&	-1.91 & --\\
\hline
\end{tabular}
\end{table*}

\begin{table*}
\caption{Normalized disc winds parameters for our sample. Note :(1) Observation number. (2) Logarithm of the minimum mean distance of the wind normalized to the Schwarschild radius. (3) Logarithm of the mean wind kinetic power normalized to the Eddington luminosity.(4) Logarithm of the ratio between the mean mass outflow rate and the accretion rate. (5) Logarithm of the ratio between the mean mass outflow rate and the Eddington accretion rate. (6)Logarithm of the ratio between the mean momentum rate of the outflow and the momentum rate of the radiation.}
\centering
\setlength{\tabcolsep}{2pt}
\begin{tabular}{lccccccc}
\hline\hline
(1)	& (2) & (3)	& (4) & (5) & (6)\\
\noalign{\smallskip}
\hline
\noalign{\smallskip}
num		& $log(\frac{r_{med}}{r_S})$	&$log(\frac{L_{med,out}}{L_{Edd}})$	  	&$log(\frac{\dot{M}_{med,out}}{\dot{M}_{acc}})$	 &$log(\frac{\dot{M}_{med,out}}{\dot{M}_{Edd}})$		 &$log(\frac{\dot{p}_{med,out}}{\dot{p}_{rad}})$	\\
\noalign{\smallskip}
\hline
\noalign{\smallskip}
1 & 3.35	$\pm$	1.27	&	-2.15	$\pm$	1.27	&	0.65	$\pm$	1.27	&	-0.77	$\pm$	1.27	&	0.61	$\pm$	1.27\\
2 & 3.95	$\pm$	1.71	&	-1.95	$\pm$	1.71	&	0.07	$\pm$	1.71	&	-0.41	$\pm$	1.71	&	-0.05	$\pm$	1.71\\
3 & 2.42	$\pm$	1.30	&	-0.70	$\pm$	1.30	&	0.16	$\pm$	1.30	&	-0.28	$\pm$	1.30	&	0.60	$\pm$	1.30\\
4 & 3.34	$\pm$	2.06	&	-0.87	$\pm$	2.06	&	0.47	$\pm$	2.06	&	-0.28	$\pm$	2.06	&	0.82	$\pm$	2.06\\
5 & 1.83	$\pm$	0.16	&	-2.17	$\pm$	0.16	&	0.41	$\pm$	0.16	&	-1.20	$\pm$	0.16	&	0.57	$\pm$	0.16\\
6 & 2.90	$\pm$	0.37	&	-2.32	$\pm$	0.37	&	1.04	$\pm$	0.37	&	-0.49	$\pm$	0.37	&	0.78	$\pm$	0.37\\
7 & 4.60	$\pm$	1.66	&	-1.12	$\pm$	1.66	&	2.43	$\pm$	1.66	&	1.12	$\pm$	1.66	&	1.96	$\pm$	1.66\\
8 & 3.64	$\pm$	1.13	&	-2.80	$\pm$	1.13	&	-0.18	$\pm$	1.13	&	-1.00	$\pm$	1.13	&	-0.43	$\pm$	1.13\\
9 & 4.00	$\pm$	2.27	&	-1.67	$\pm$	2.27	&	0.39	$\pm$	2.27	&	-0.65	$\pm$	2.27	&	0.53	$\pm$	2.27\\
10 & 1.96	$\pm$	0.65	&	-1.19	$\pm$	0.65	&	0.17	$\pm$	0.65	&	-0.58	$\pm$	0.65	&	0.51	$\pm$	0.65\\
11 & 3.95	$\pm$	1.46	&	-1.86	$\pm$	1.46	&	0.08	$\pm$	1.46	&	-0.07	$\pm$	1.46	&	-0.17	$\pm$	1.46\\
12 & 3.76	$\pm$	0.58	&	-3.24	$\pm$	0.58	&	0.64	$\pm$	0.58	&	-0.76	$\pm$	0.58	&	0.05	$\pm$	0.58\\
13 & 3.26	$\pm$	1.02	&	-2.68	$\pm$	1.02	&	0.24	$\pm$	1.02	&	-1.14	$\pm$	1.02	&	0.13	$\pm$	1.02\\
14 & 1.95	$\pm$	0.74	&	-1.26	$\pm$	0.74	&	0.58	$\pm$	0.74	&	-0.74	$\pm$	0.74	&	0.97	$\pm$	0.74\\
15 & 2.25	$\pm$	1.12	&	-1.02	$\pm$	1.12	&	0.62	$\pm$	1.12	&	-0.60	$\pm$	1.12	&	1.05	$\pm$	1.12\\
16 & 2.67	$\pm$	1.40	&	-1.02	$\pm$	1.40	&	0.33	$\pm$	1.40	&	-0.45	$\pm$	1.40	&	0.69	$\pm$	1.40\\
17 & 2.45	$\pm$	0.44	&	-2.85	$\pm$	0.44	&	-0.32	$\pm$	0.44	&	-1.54	$\pm$	0.44	&	-0.33	$\pm$	0.44\\
18 & 2.02	$\pm$	0.02	&	-2.40	$\pm$	0.02	&	0.57	$\pm$	0.02	&	-1.10	$\pm$	0.02	&	0.57	$\pm$	0.02\\
19 & 2.76	$\pm$	1.68	&	-1.60	$\pm$	1.68	&	-0.28	$\pm$	1.68	&	-1.22	$\pm$	1.68	&	0.17	$\pm$	1.68\\
20 & 2.66	$\pm$	0.59	&	-1.74	$\pm$	0.59	&	0.88	$\pm$	0.59	&	-0.37	$\pm$	0.59	&	0.84	$\pm$	0.59\\
21 & 3.62	$\pm$	1.10	&	-2.74	$\pm$	1.10	&	-0.85	$\pm$	1.10	&	-0.92	$\pm$	1.10	&	-1.11	$\pm$	1.10\\
22 & 2.90	$\pm$	0.70	&	-1.75	$\pm$	0.70	&	0.52	$\pm$	0.70	&	-0.24	$\pm$	0.70	&	0.42	$\pm$	0.70\\
23 & 2.34	$\pm$	0.77	&	-1.19	$\pm$	0.77	&	0.93	$\pm$	0.77	&	-0.31	$\pm$	0.77	&	1.14	$\pm$	0.77\\
24 & 3.25	$\pm$	1.61	&	-0.89	$\pm$	1.61	&	0.55	$\pm$	1.61	&	0.05	$\pm$	1.61	&	0.73	$\pm$	1.61\\
25 & 2.68	$\pm$	0.81	&	-2.10	$\pm$	0.81	&	0.47	$\pm$	0.81	&	-0.93	$\pm$	0.81	&	0.54	$\pm$	0.81\\
26 & 3.27	$\pm$	1.87	&	-1.46	$\pm$	1.87	&	-0.64	$\pm$	1.87	&	-0.75	$\pm$	1.87	&	-0.34	$\pm$	1.87\\
27 & 2.58	$\pm$	1.49	&	-0.27	$\pm$	1.49	&	2.04	$\pm$	1.49	&	0.12	$\pm$	1.49	&	2.49	$\pm$	1.49\\

\hline
\end{tabular}
\end{table*}

\bsp	
\label{lastpage}
\end{document}